\documentclass[7pt, a4paper, twoside, twocolumn]{article}
\usepackage{amsfonts}
\usepackage{a4}
\usepackage{graphicx}
\usepackage{amsmath,amssymb}
\usepackage{graphicx}
\usepackage{color}
\usepackage{multirow}
\usepackage[comma, super]{natbib}
\makeatother

\date{}

\begin{document}
\definecolor{grund}{gray}{.5}
\definecolor{orange}{rgb}{0.255, 0.165, 0}
\definecolor{dg}{gray}{0.25}
%
\begin{titlepage}
\sf
  \noindent{\Huge Bag-of-calls analysis reveals group-specific vocal repertoire in long-finned pilot whales}
  \\[0.5cm]
\noindent{
\large
Heike Vester$^{1,2,3\ast}$,  
Kurt Hammerschmidt$^{3}$,
Marc Timme$^{2,4}$,
Sarah Hallerberg$^{2,4}$
}
\\[1.5cm]
%
%
\noindent{
\large 
Besides humans, several marine mammal species exhibit prerequisites to evolve language: high cognitive abilities, flexibility in vocal production and advanced social interactions. 
Here, we describe and analyse the vocal repertoire of long-finned pilot whales (Globicephalus melas) recorded in northern Norway. 
Observer based analysis reveals a complex vocal repertoire with 140 different call types, call sequences, call repetitions and group-specific differences in the usage of call types. 
Developing and applying a new automated analysis method, the bag-of-calls approach, we find that groups of pilot whales can be distinguished purely by statistical properties of their vocalisations. 
Comparing inter- and intra-group differences of ensembles of calls allows to identify and quantify group-specificity. 
Consequently, the bag-of-calls approach is a valid method to specify difference and concordance in acoustic communication in the absence of exact knowledge about signalers, which is common observing marine mammals under natural conditions.
}
\vspace{8cm}
{\bf
\textcolor{orange}{\hrule}
}
\vspace{0.2cm}
\noindent{
\small
{\sf $^{1}$Ocean Sounds, Sauoya 01, 8312 Henningsvaer, Norway};
{\sf $^{2}$Network Dynamics, Max Planck Institute for Dynamics and Self-Organization, 37077 G{\"o}ttingen, Germany}; {$^{3}$Cognitive Ethology Lab, German Primate Center, Kellnerweg 4, 37077 G{\"o}ttingen, Germany}; {Institute for Nonlinear Dynamics, Faculty for Physics, 37077 G{\"o}ttingen, Germany}
}
\\[0.2cm]
\end{titlepage}
%
%
\vspace{-0.2cm} 
According to the social function hypothesis, the vocal repertoire of a species should become more diverse the more complex its social system is \citep{Freeberg2012, McComb2005}. 
In societies of social whales group composition is stable over many years or generations and maternal care is long, sometimes lasting for a lifetime. 
Complex, vocalisations are characteristic for matrilineal social whales, and are known to be often aquired by learning, as e.g. in killer whales ({\sl Orcinus orca})\citep{Deecke2000,Yurk2002, Rendell2003}). 
%
%
The existence of vocal cultures and dialects has been suggested by observer based analysis of variations in call type usage of killer whales \cite{Ford1991, Deecke2010} and sperm whale codas \cite{Rendell2012}. 
In contrast, more solitary baleen whales such as blue whales ({\sl Balaenoptera musculus})(see e.g.~\citep{Berchok2006}) or fin whales ({\sl Balaenoptera physalus}) \citep{Edds1988} use fewer vocalisations with some geographical differences. 
Less is known about the vocalisations of long-finned pilot whales, especially about the polulation living in the northeast atlantic.

Long-finned pilot whales ({\sl Globicephalus melas}) belong to the dolphin family and represents the second largest dolphin species after the killer whales, with adult males reaching 6.5 m and females 5.5 m in length \citep{Bloch1993}. 
The species is widely distributed and is found in circumpolar regions both in the northern and southern hemispheres and in the Mediterranean (see \citep{Rice1998} and refs.~therein). 
Long-finned pilot whales mainly occur in waters deeper than 100 m, often at the edge of a geographical drop off, with migrations between offshore and inshore waters, correlating with the distribution of squid, their main prey. 
In the North Atlantic they mainly feed on squid {\sl Gonatus spp} and {\sl Todarodes sagittatus}, but occasionally they may feed on fish as well \citep{Desportes1993}.

Long-finned pilot whales in the {\sl northwest} atlantic produce typical dolphin sounds, such as clicks, buzzes, grunts, and a variety of pulsed calls including whistles \citep{Weilgart1990, Nemiroff2009}. 
Pulsed calls of long-finned pilot whales in the northwest atlantic population are similar in structure to killer whale calls. 
They are complex with different structural components, such as elements and segments, and a fifth of the calls can be bi-phonal with a lower (LFC) and an upper frequency component (UFC) \citep{Yurk2005,Nemiroff2009}. 
In contrast to discrete killer whale call types, pilot whales in the northwest atlantic have been reported to use graded calls \citep{Nemiroff2009}. 

In this contribution, we find very different results for the previously unknown vocal repertoire of the long-finned pilot whale population in the {\sl northeast atlantic}, i.e., in northern Norway.
An observer based analysis reveals the existence of a complex vocal repertoire with 140 discrete call types, call sequences, call repetitions and group-specific differences in the usage of call types.

Introducing a new automated method to study group-specificity, the {\sl bag-of-calls approach} (BOC) we quantify group-specific differences in vocalisation.
Previous approaches to automated analysis aimed on automated categorisation of calls \citep{Spong1993,Deecke1999, Deecke2005, Brown2007, Brown2009, Kaufman2012}. 
We however, test whether one can study the communication of whales without categorising and classifying {\sl single calls} and instead propose and apply an automated analysis method to collections ({\sl ensembles}, also called {\sl bags}) of calls or recordings. 
Computing distributions of cepstral coefficients \citep{Bogert1963} for each ensembles, allow us to quantify group-specificity in a statistical significant way.
%
%
\section*{\sf \textbf{Results}}
\begin{figure}[h!]
\centerline{
\includegraphics[width=0.5\textwidth]{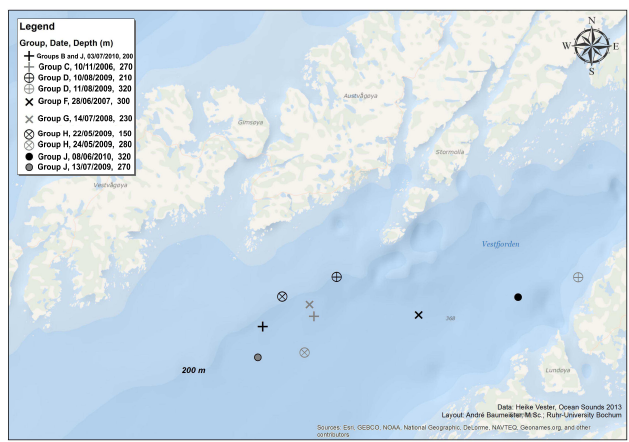}
}
\caption{\label{map}
\sf \textbf{Map of encounters with long-finned pilot whale groups in the Vestfjord (Northern Norway).} Symbols mark the places where the different groups have been encountered. More details about the different encounters can be found in Tab.~\ref{tab:encounters}.
}
\end{figure}

{\sf \textbf{Vocal Repertoire.}}
Studying six groups of long-finned pilot whales in the Vestfjord, in northern Norway (see Fig.~\ref{map} and Tab.~\ref{tab:encounters})
\begin{table*}[t!]
\sf
\caption{\sf \textbf{Encounters, recordings and behaviour.}}
\vspace{0.2cm}
\begin{tabular}{c c c c c c c}
\hline
& & & &  & \textbf{sound rec.} &\\
 & & & & & \textbf{[h:min]} & \\ 
& & & & &  and &\\
 & & & & \textbf{obs.~} & \textbf{number of} & \\
& \textbf{date and location} & \textbf{estimated} & \textbf{photo} &  \textbf{time} & \textbf{recording} & \\
\textbf{group} & \textbf{[ N/E ]} & \textbf{group size} & \textbf{ID's } & \textbf{[h:min]} & \textbf{sessions} & \textbf{behaviour}\\[0.1cm]
\hline
{\multirow{2}{*}{B}} & 03/07/2010 & 45 & 43 & 05:33 & 03:17 & slow travelling,\\
& 68$^{\circ}$04.870$^{\prime}$/14$^{\circ}$25.130$^{\prime}$ & & & & 19 & socialising \\
\hline
{\multirow{8}{*}{D}} & 10/08/2009 & 100 & 60 & 04:42 & 01:07 & milling,\\
& 68$^{\circ}$06.532$^{\prime}$/14$^{\circ}$32.771$^{\prime}$ & & & & 7 & slow travelling, \\
& & & & & & socialising \\
\cline{2-7}
  & 11/08/2009 & 100 & 60 & 04:11 & 02:25 & milling,\\
& 68$^{\circ}$10.997$^{\prime}$/15$^{\circ}$29.205$^{\prime}$ & & & & 15 & slow travelling, \\
& & & & & & socialising, \\
& & & & & & foraging, \\
& & & & & & resting \\
\hline
{\multirow{2}{*}{F}} &  28/06/2007 & 20 & 9 & 00:50 & 00:23 &  milling,\\ 
& 68$^{\circ}$04.517$^{\prime}$/14$^{\circ}$49.436$^{\prime}$ & & & & 3 & boat friendly\\
\hline
{\multirow{3}{*}{G}} & 14/07/2008 & 7 & 4 & 02:10 & 00:49 & milling,\\
& 68$^{\circ}$07.612$^{\prime}$/14$^{\circ}$40.562$^{\prime}$ & & & & 7 & socialising, \\
& & & & & & boat friendly \\
\hline
{\multirow{5}{*}{H}} & 22/05/2009 & 50 & 32 & 04:00 & 02:45 & milling,\\
& 68$^{\circ}$08.578$^{\prime}$/14$^{\circ}$31.581$^{\prime}$ & & & & 19 & socialising, \\
& & & & & & resting \\
\cline{2-7}
& 24/05/2009 & 50 & 22 & 02:00 & 01:20 & milling,\\
& 68$^{\circ}$01.636$^{\prime}$/14$^{\circ}$38.966$^{\prime}$ & & & & 9 & resting \\
\hline
{\multirow{13}{*}{J}} & 13/07/2009 & 60 & 17 & 05:48 & 03:10 & milling, \\
& 68$^{\circ}$01.053$^{\prime}$/14$^{\circ}$23.519$^{\prime}$ & & & & 19 & resting \\
& & & & & & socialising, \\
& & & & & & boat friendly \\
\cline{2-7}
& 08/06/2010 & 60 & 19 & 01:40 & 01:06 & first fast\\
& 68$^{\circ}$08.540$^{\prime}$/15$^{\circ}$09.330$^{\prime}$ & & & & 9 & travelling\\
& & & & & & -avoided boat, \\
& & & & & & later \\
& & & & & & calmed down \\
& & & & & & slow travelling, \\
\cline{2-7}
& 03/07/2010 & n/a & 4 & 02:00 & 01:08 & travelling\\
& 68$^{\circ}$04.870$^{\prime}$/14$^{\circ}$25.130$^{\prime}$ & & & & 3 &  and milling \\
& & & & & & with group B \\
\hline
\end{tabular}
\begin{flushleft}
{\sf
Group sizes are estimated by visual observation on site, whereas the number of photo ID's refer to animals identified a-posteriori from pictures taken.
During each encounter several recordings were made (number of recording sessions). Summing the duration of these recording sessions yields the total duration of recordings presented here. 
}
\end{flushleft}
\label{tab:encounters}
\end{table*}
we find a complex and flexible vocal repertoire, containing more than 140 different call-types.
Using a total of 32:54 h of observation time and   
17:32 h of sound recordings in 99 recording sessions, 4582 calls could be categorised in different types using classic audio-visual observer classification. 
\begin{figure}[t!]
\centerline{
\includegraphics[width=0.5\textwidth]{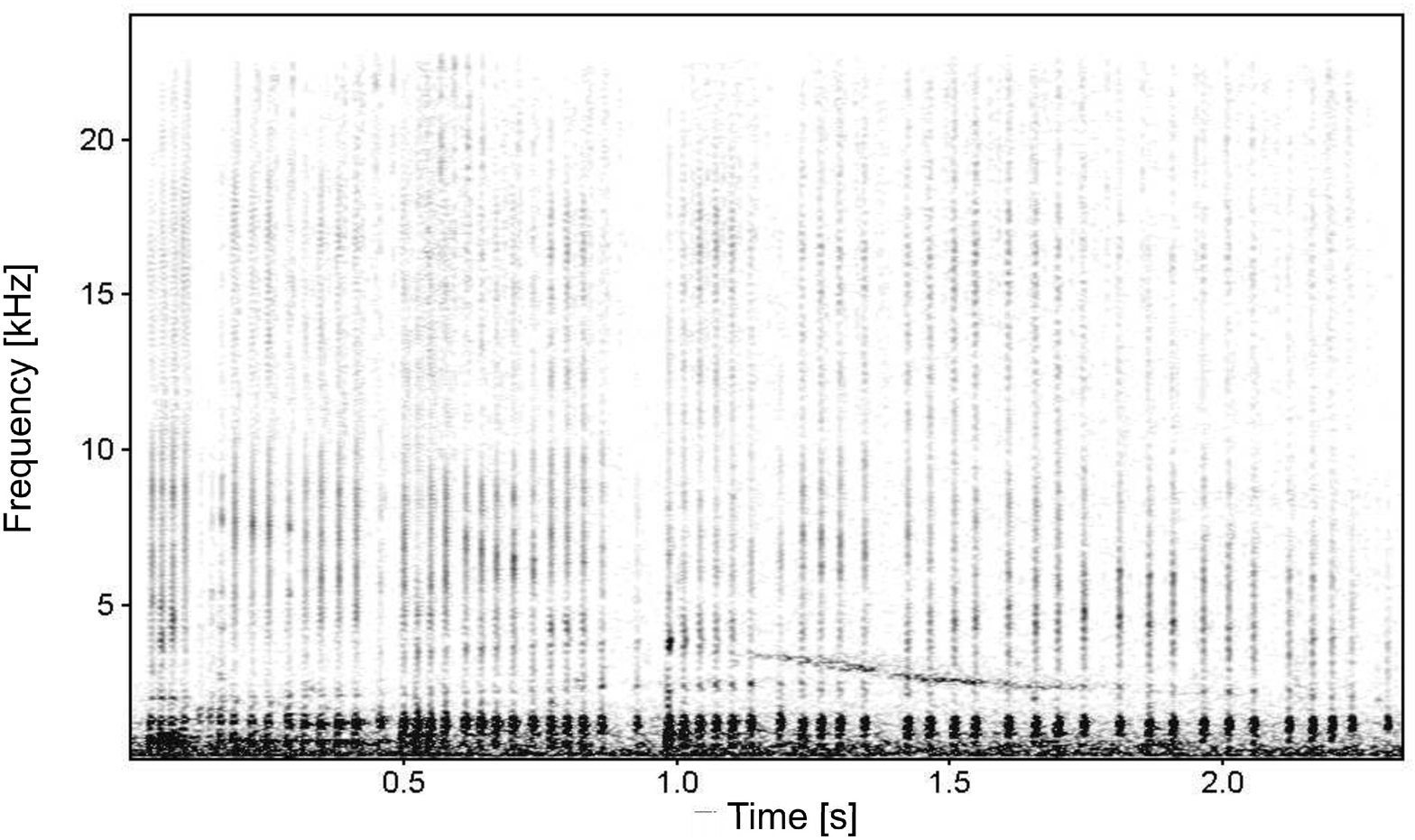}
\put(-200,120){\sf (a)}
}
\centerline{
\includegraphics[width=0.5\textwidth]{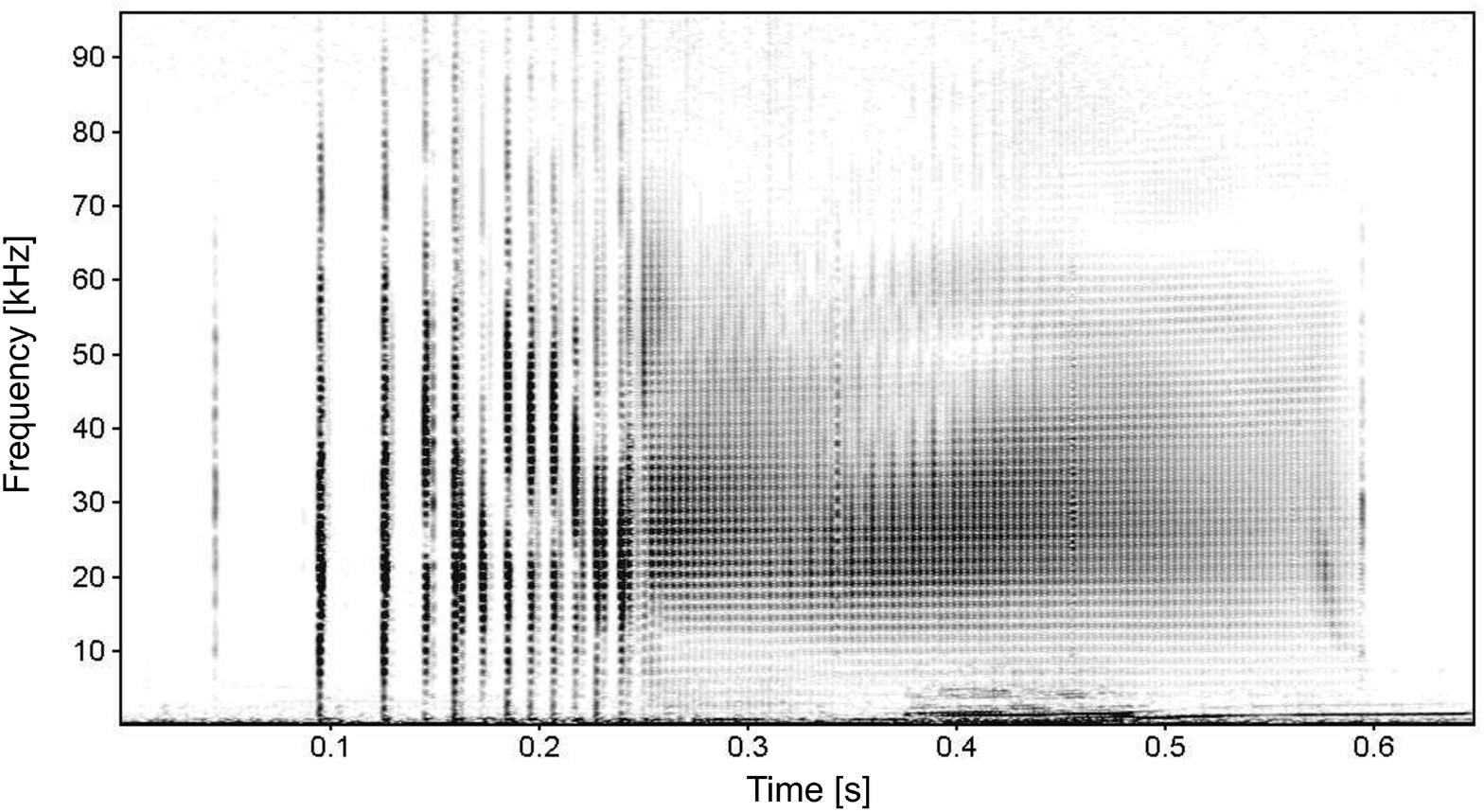}
\put(-210,110){\sf (b)}
}
\centerline{
\includegraphics[width=0.5\textwidth]{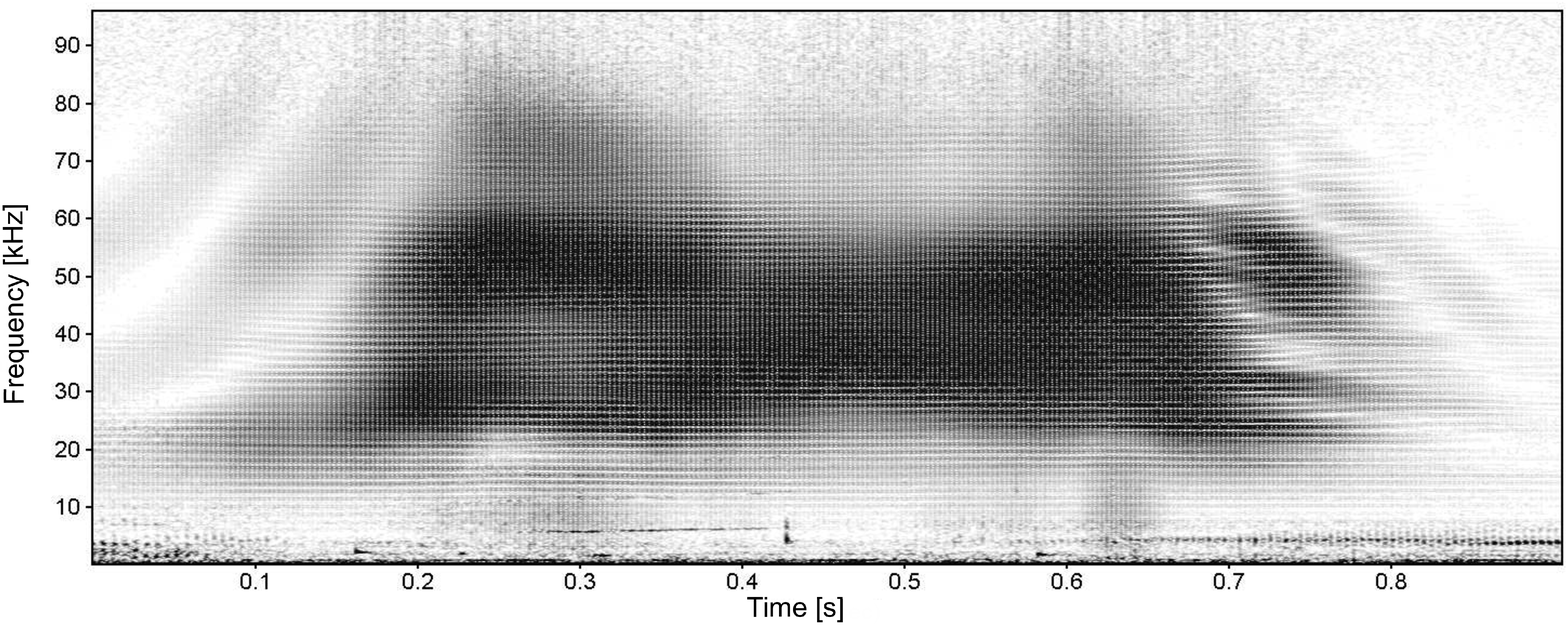}
\put(-210,75){\sf (c)}
}
\caption{\label{specs1} \sf \textbf{Spectrograms of examples of typical long-finned pilot whale sounds recorded in northern Norway}: 
Examples of different clicks, buzzes, and variable or non-stereotyped calls.
(a) Low frequency clicks, most power below 2 kHz.
(b) High frequency clicks and buzzes, most power above 20 kHz.
(c) High frequency buzzes with frequencies 20-60 kHz.
}
\end{figure}
\begin{figure}[ht!!]
\centerline{
\includegraphics[width=0.22\textwidth, height=6.8cm]{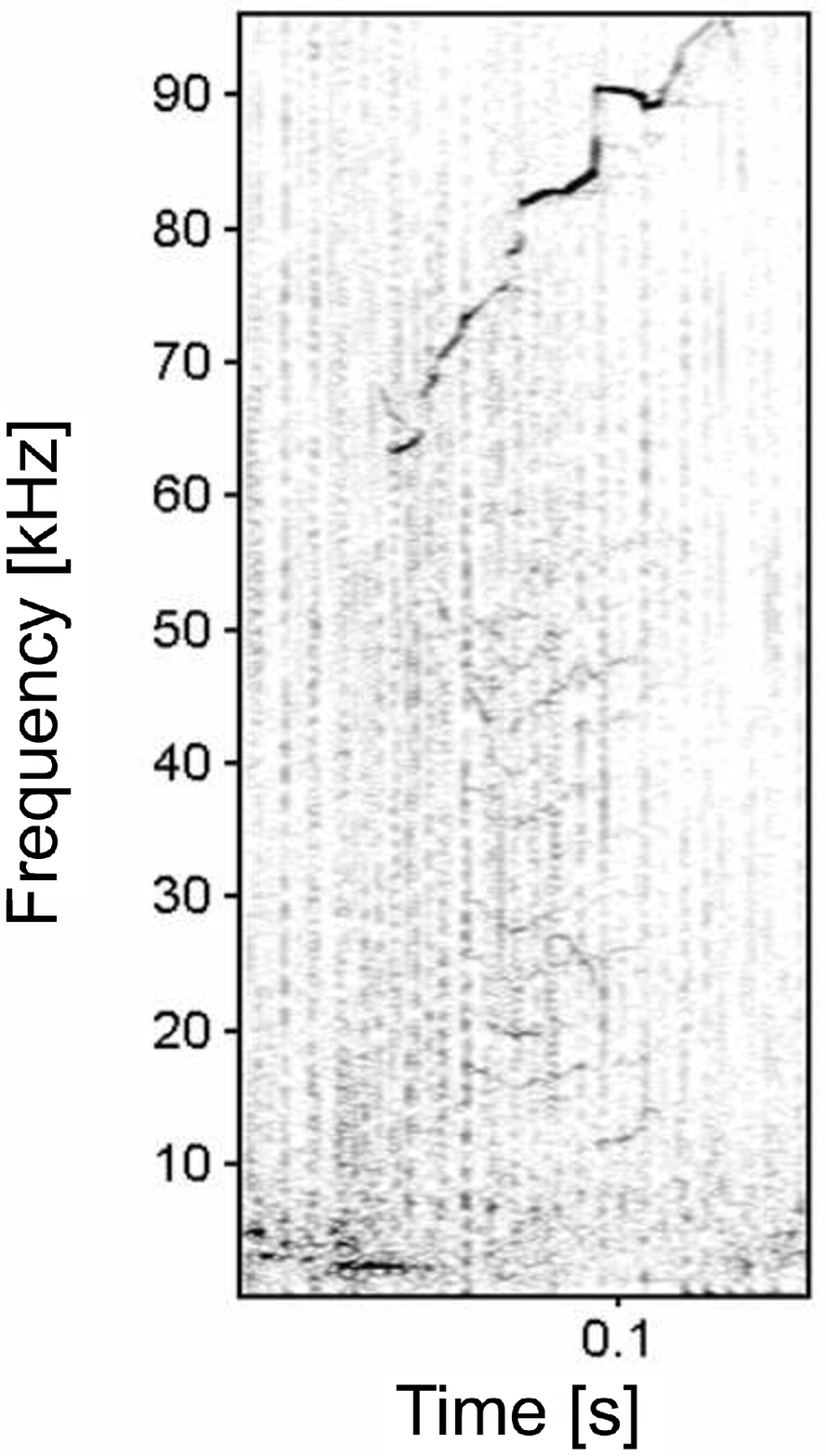}
\put(-70,172){\sf (a)}
\hspace{-3mm}
\includegraphics[width=0.28\textwidth, height=7.2cm]{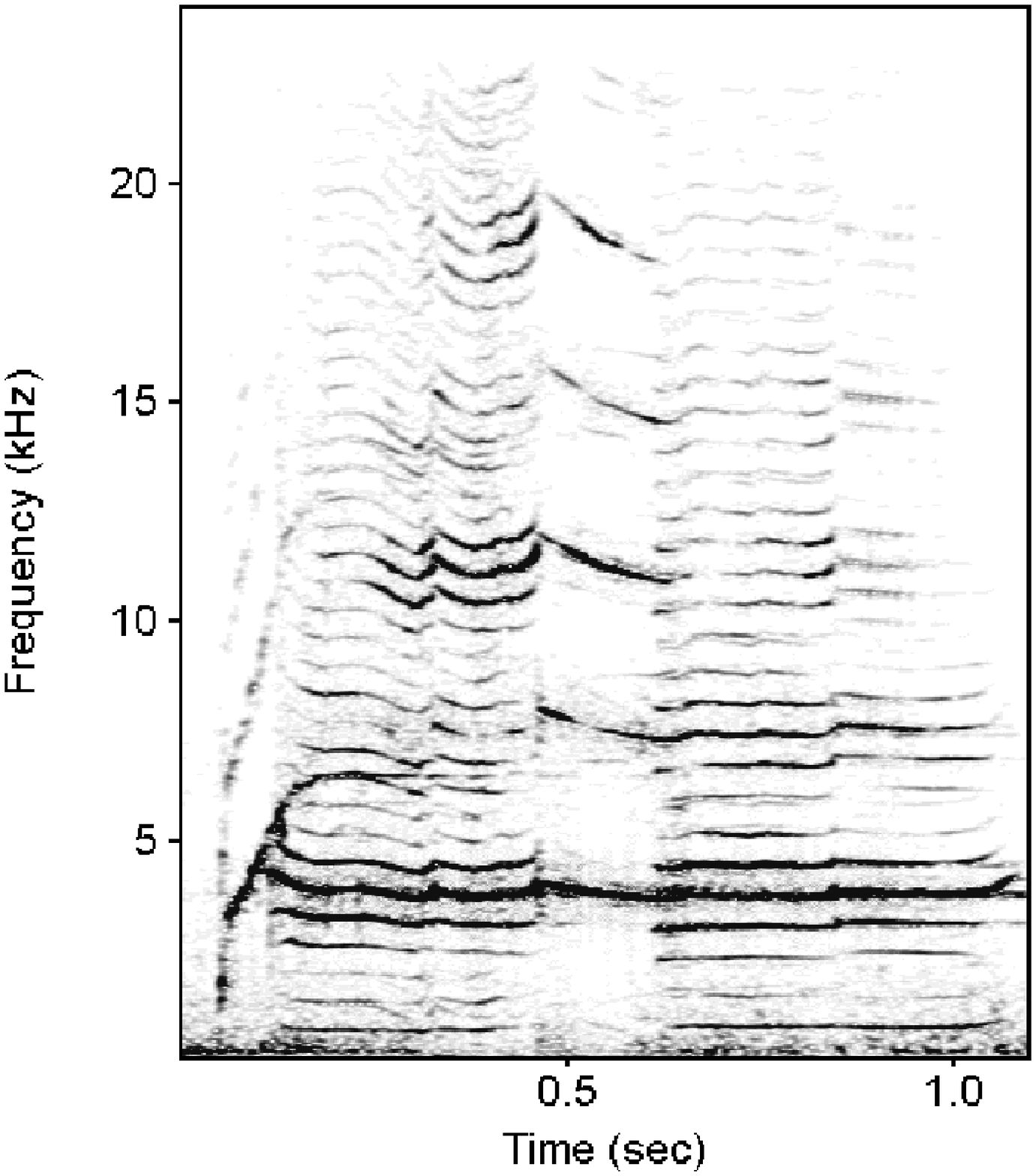}
\put(-100,190){\sf (b)}
}
\centerline{
\includegraphics[width=0.51\textwidth]{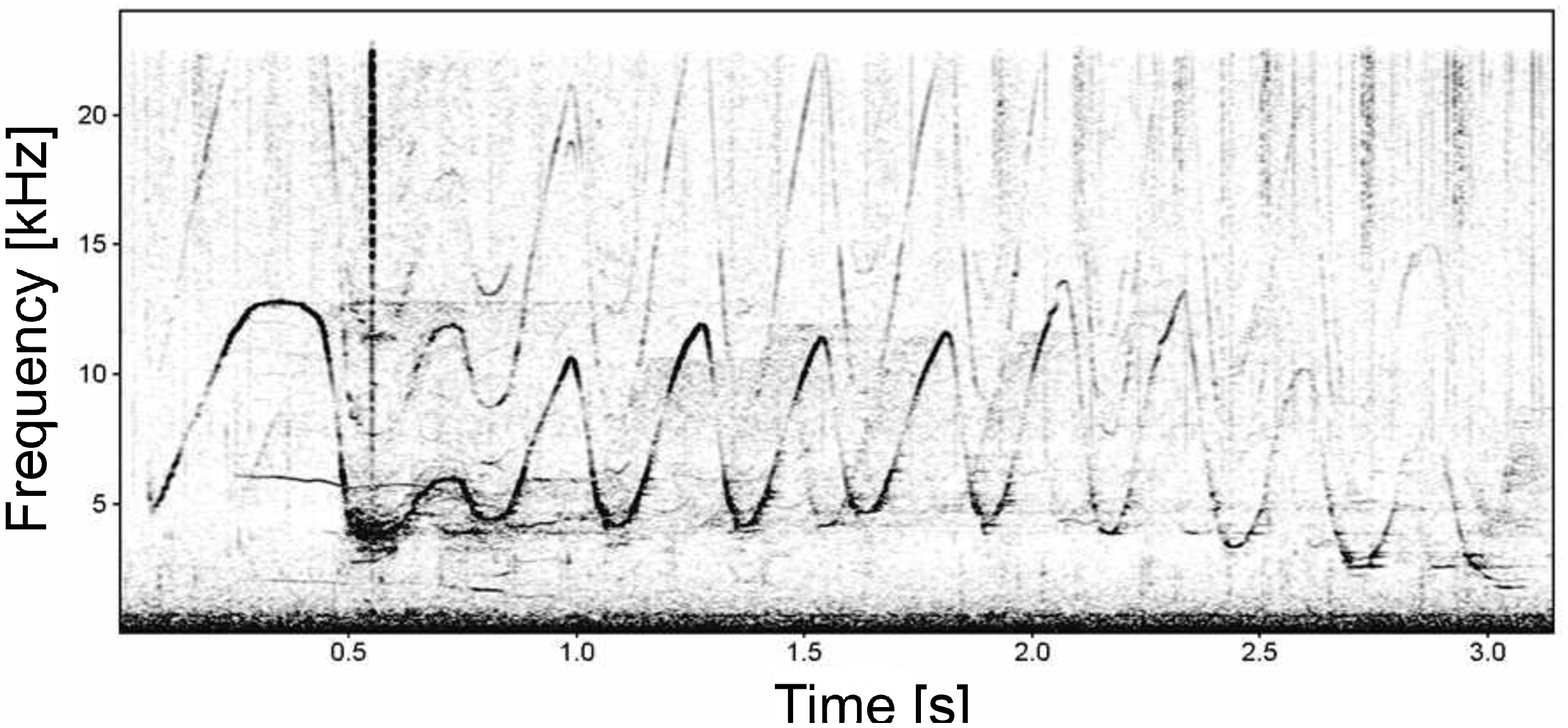}
\put(-220,90){\sf (c)}
}
\centerline{
\includegraphics[width=0.5\textwidth]{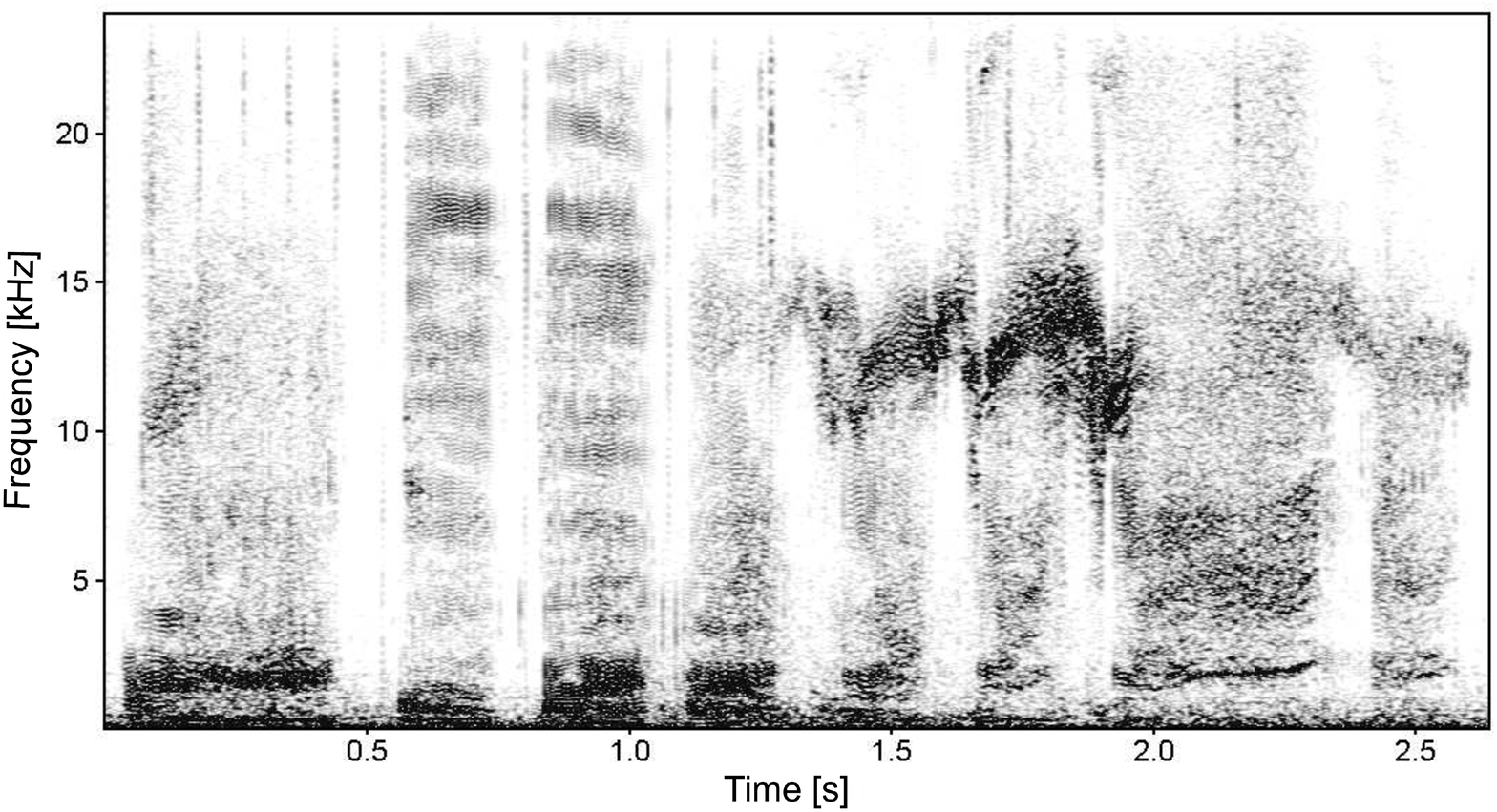}
\put(-210,108){\sf (d)}
}  
\caption{\label{specs2} \sf \textbf{Examples of different pulsed calls.} 
(a) High frequency (HF) whistles above 60 kHz (from one group H; N=6).
(b) Stereo-typed call of one segment, 6 elements and multiple frequency components.
(c) Lower frequency (LF) whistles below 20 kHz, all groups produced LF whistles.
(d) Noisy calls – low frequency sounds with no distinct frequency band often produced in sequences (“rasps” or “pig” like sounds).
}
\end{figure}
The vocal repertoire of long-finned pilot whales in northern Norway includes a variety of clicks and buzzes (see Fig.~\ref{specs1}), low frequency calls that are noisy and irregular (see Fig.~\ref{specs2}), different types of pulsed calls that range from simple and short to highly complex structures, and three types of whistles, which are below and {\sl above 20 kHz}.
Ultrasound whistles (above 20 kHz) have previously not been described for long-finned pilot whales.
%
%
\begin{figure}[ht!!!]
  \centerline{
    \hspace{-0.5cm}
\includegraphics[width=0.5\textwidth]{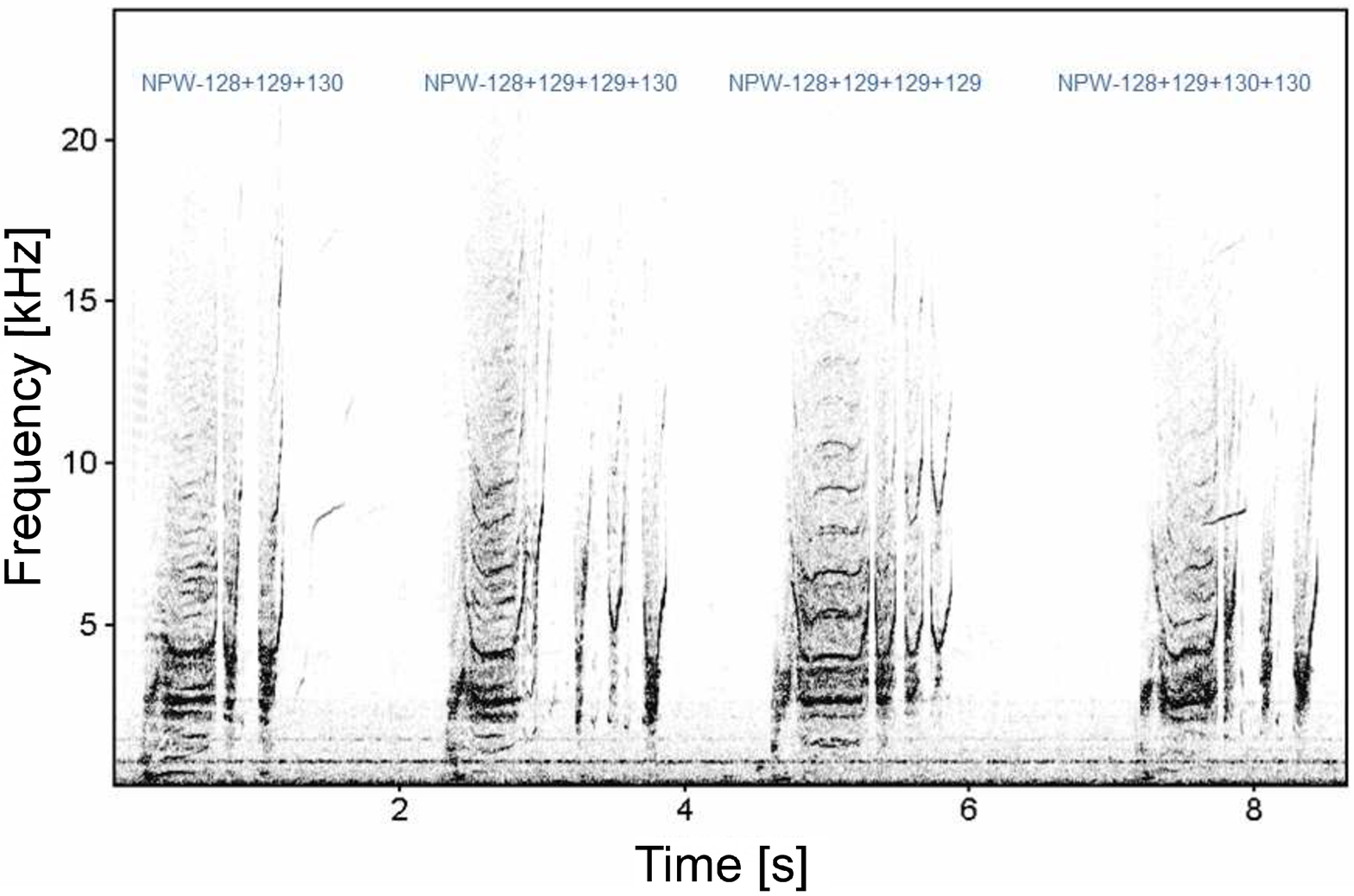}
\put(-210,120){\sf (a)}
}
\vspace{-0.5cm}
\centerline{
\includegraphics[width=0.5\textwidth]{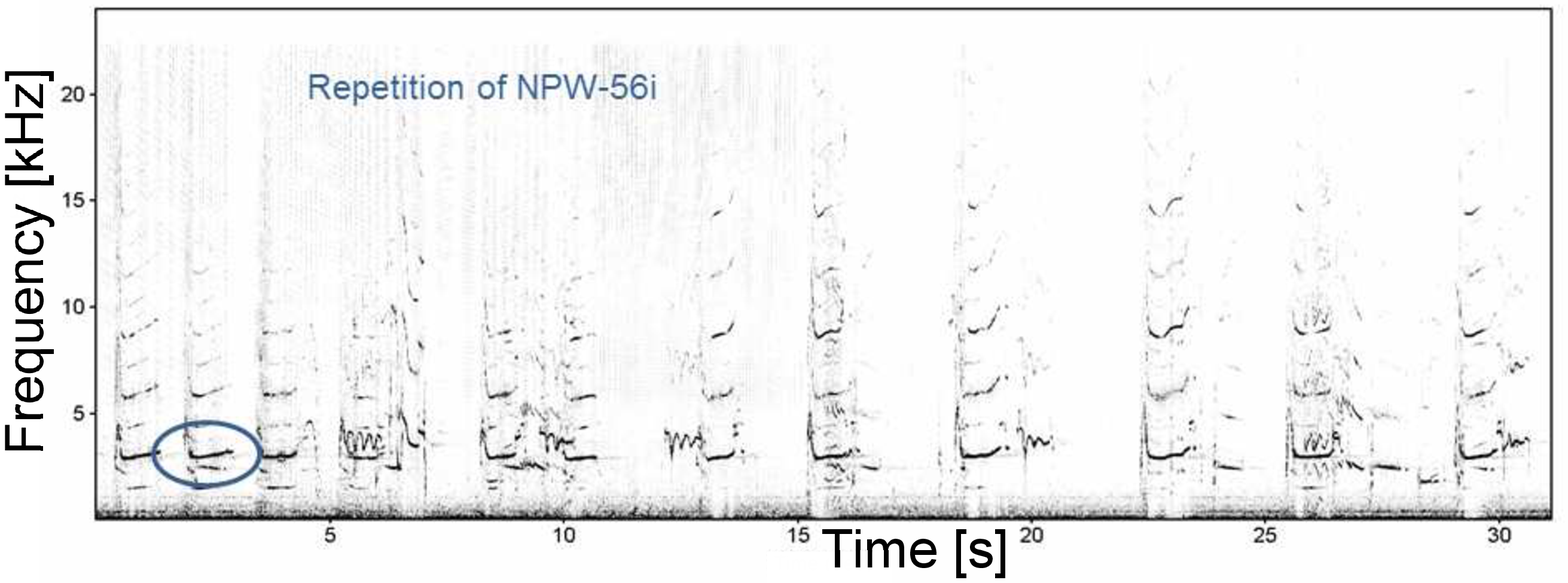}
\put(-220,75){\sf (b)}
}
\caption{\label{repetition} \sf \textbf{Combinations and repetitions of different call types.} Spectrogram of combinations (a) and repetitions (b) of different call types. The labelling in (a) refers to the respective call type numbers.}
\end{figure}
In addition, we found stereotyped call type combinations of certain call types with different variations (Fig.~\ref{repetition} (a)). 
Some call types (such as NPW-56i) were repeated more often than others. The duration of call repetitions ranged from several seconds to over $2$ minutes (Fig.~\ref{repetition} (b)).\\

\begin{figure}[ht!]
\centerline{
  \includegraphics[width=0.5\textwidth]{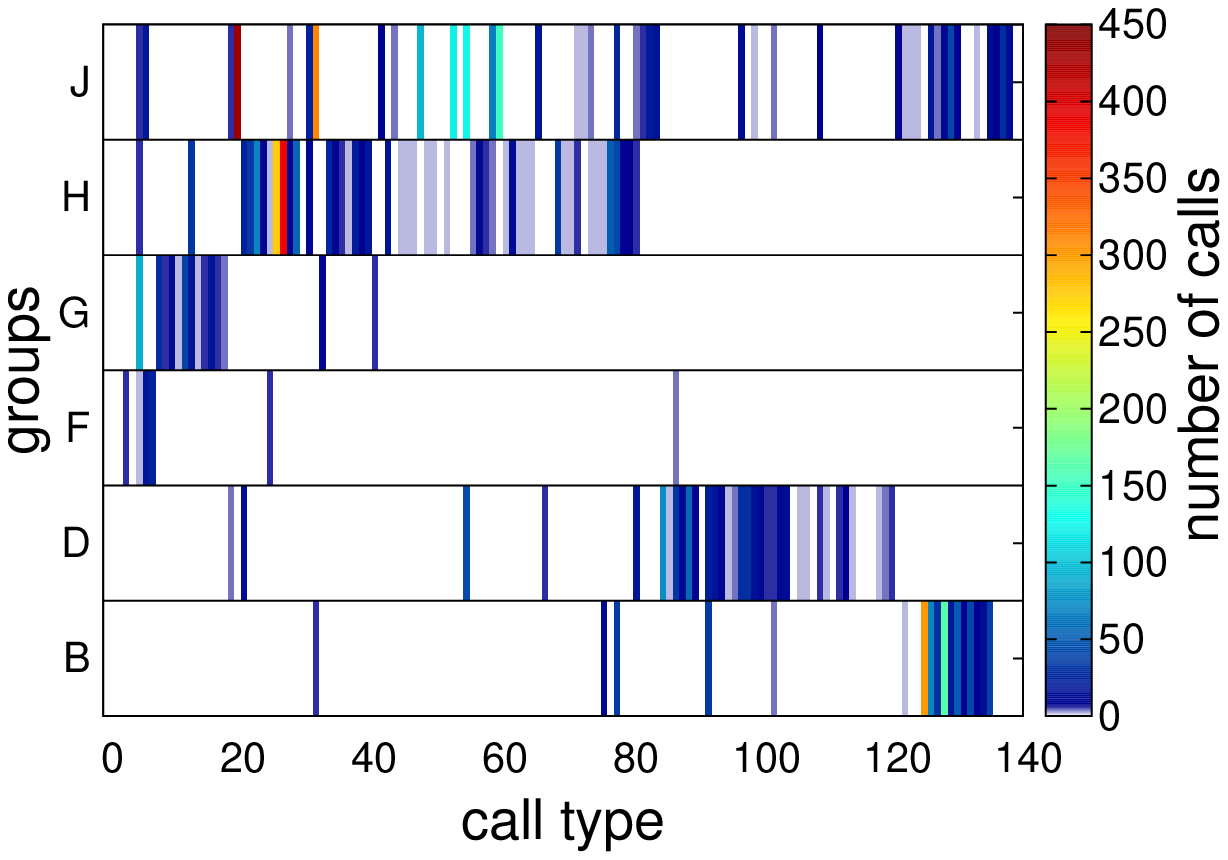}
  \put(-240, 135){\large \sf(a)}
}
\centerline{
\begin{minipage}[t!]{0.25\textwidth}
  \hspace{-1.5cm}
  \includegraphics[width=1\textwidth]{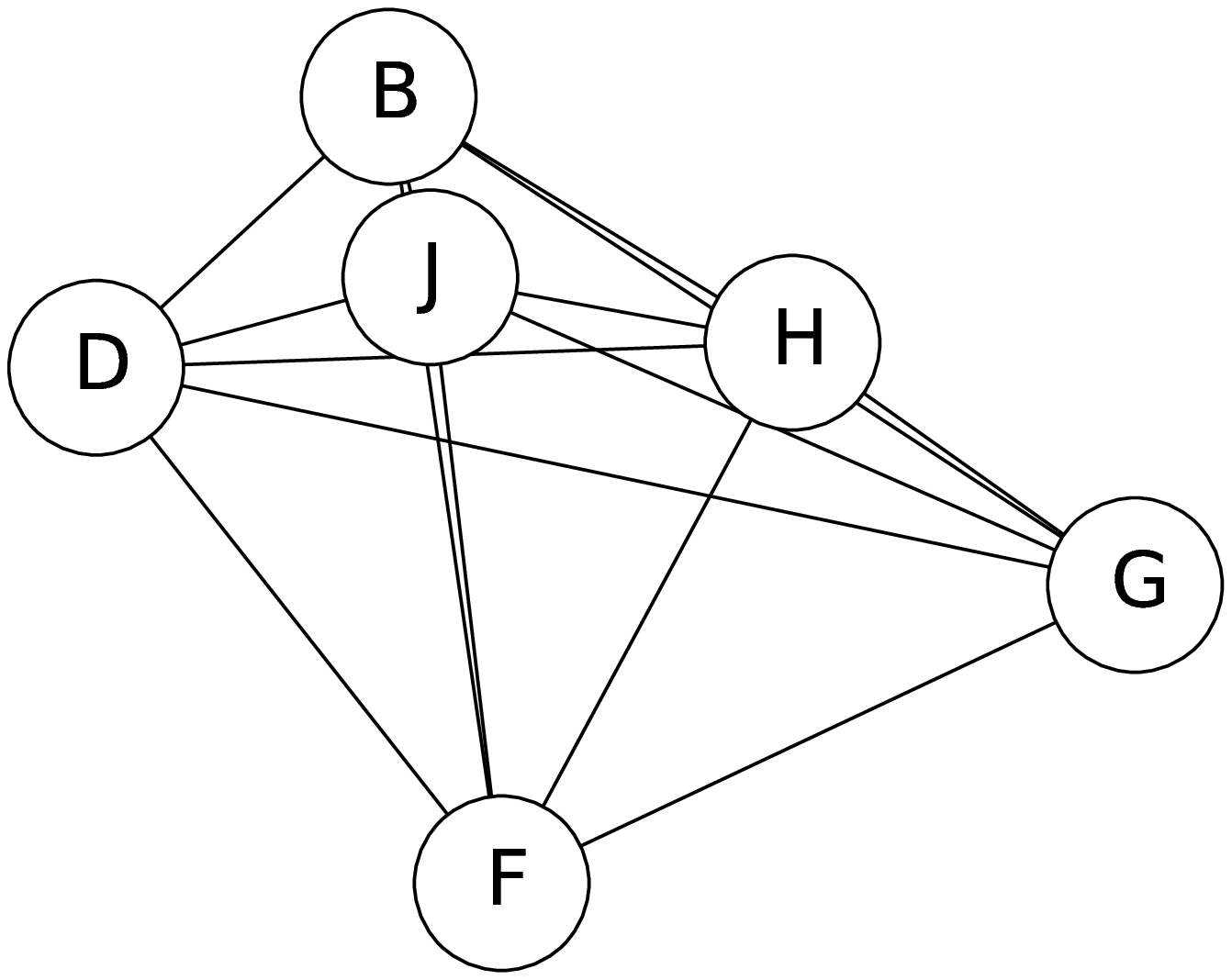} 
  \put(-100, 80){\large \sf(b)}
\end{minipage}
\hspace{-2.5cm}
\begin{minipage}[t!]{0.25\textwidth}
 \includegraphics[width=1\textwidth, angle=-90]{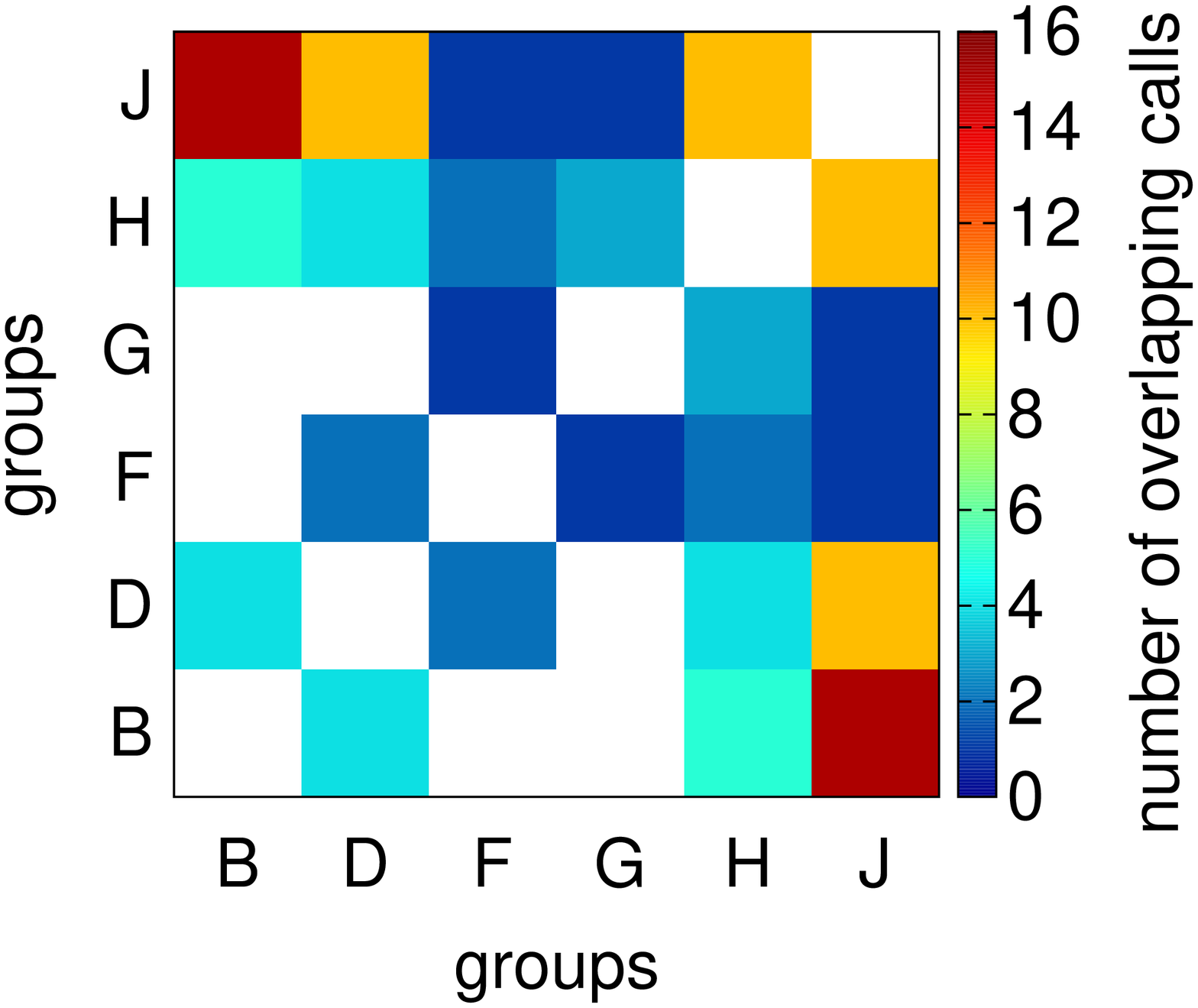}
\put(-162, -25){\large \sf(c)}
\end{minipage}
}
\caption{\label{groupvscalls} \sf \textbf{Overlap in call usage.} (a) Observations of calls (colour coded) within different groups of pilot whales. Note that call type numbering is arbitrary and does not reflect similarity in call structure, i.e., call $i$ and $i+1$ do not necessarily have to be similar in structure. Figs.~(b) and (c) show a similarity network and it's adjacency matrix estimated by the OCU approach. Here, the distance between edges in the network and the colour coding of the matrix represent the number of shared call-types.
}
\end{figure}

\begin{figure*}[ht!!!]
  \begin{minipage}[t!]{0.5\textwidth}
    \hspace{-0.5cm}
\includegraphics[width=1\textwidth]{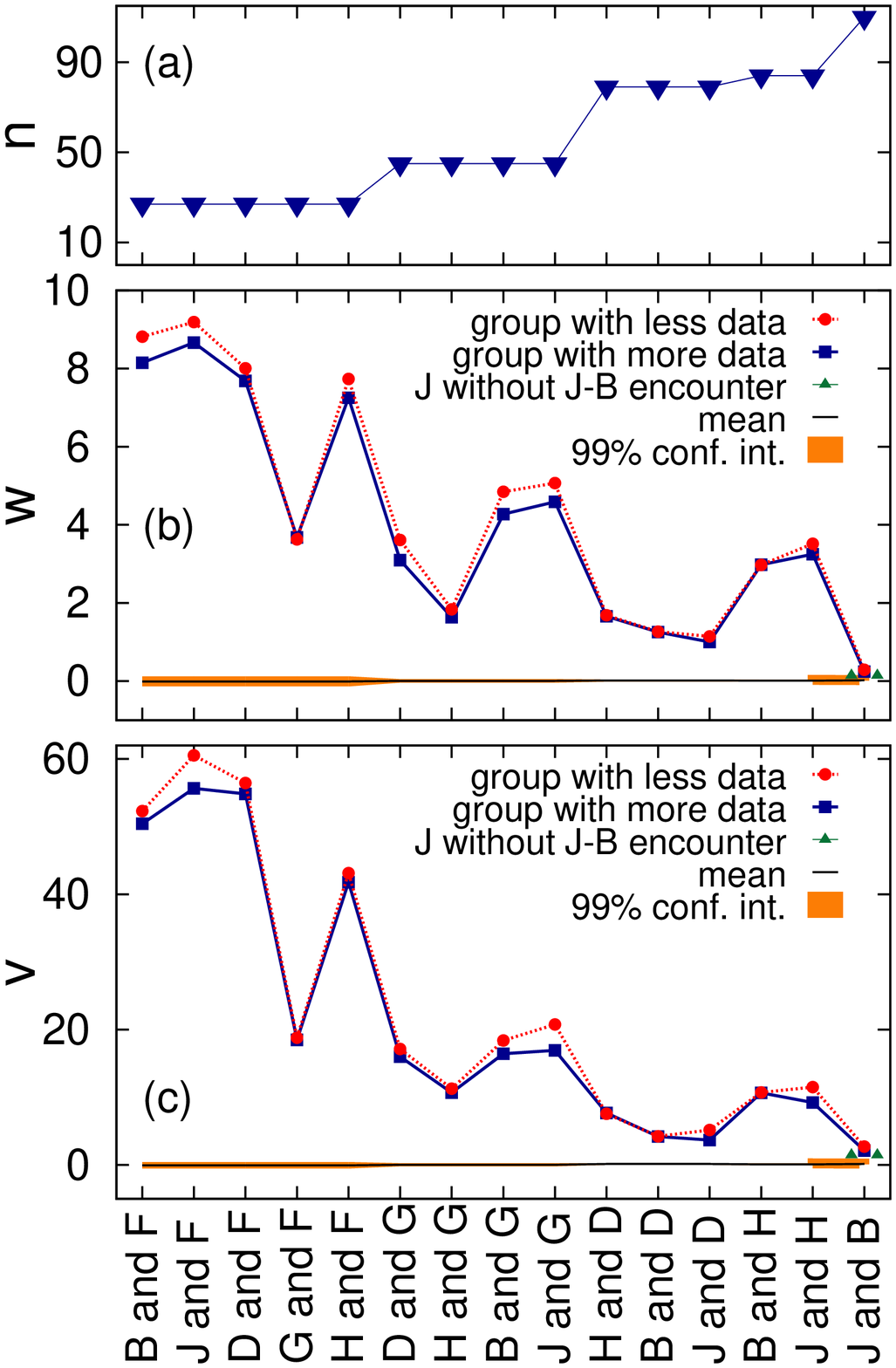}
\end{minipage}
\hspace{-0.5cm}
\begin{minipage}[t!]{0.5\textwidth}
\vspace{-1cm}
  \centerline{
    \includegraphics[width=1.1\textwidth]{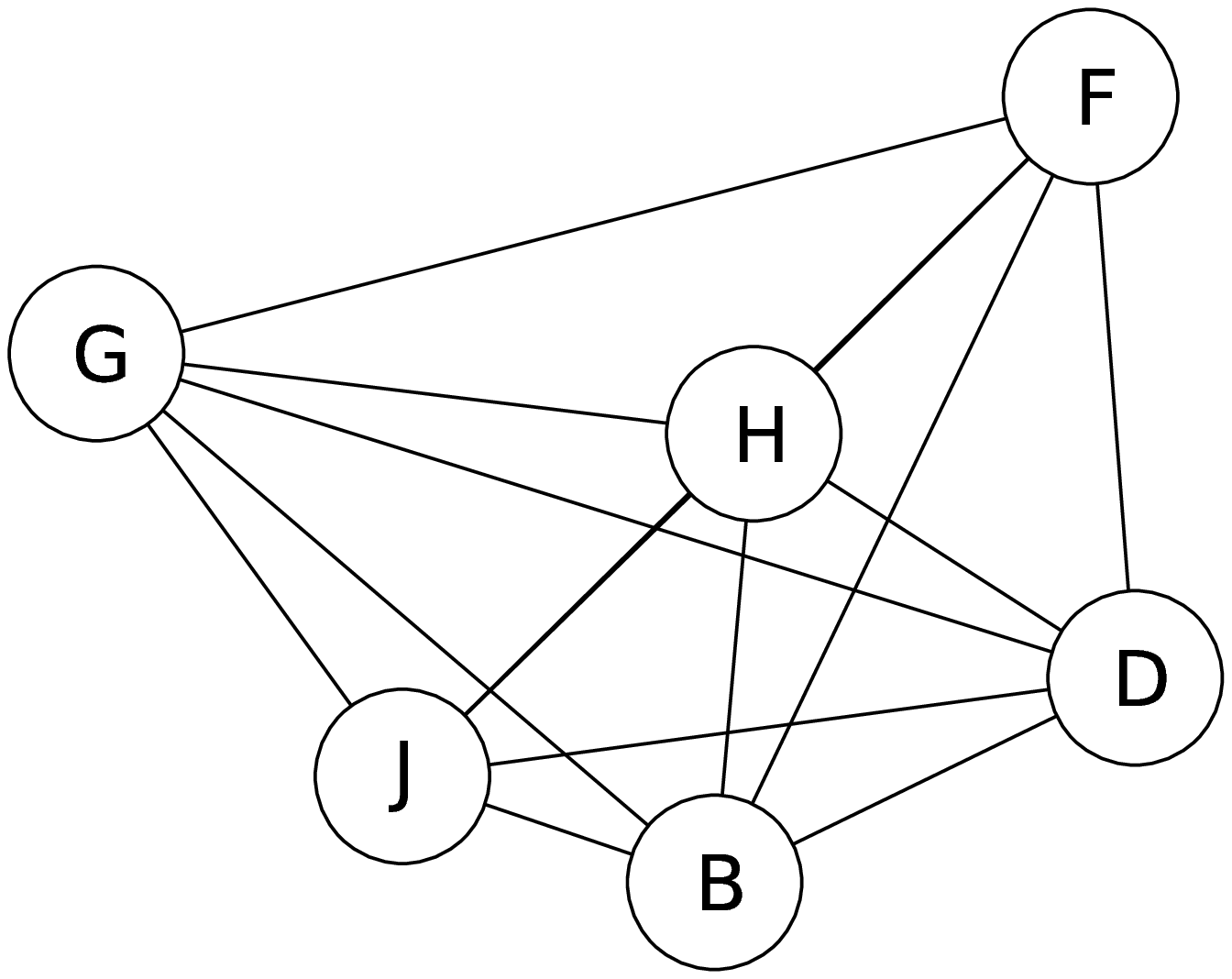}
    \put(-200, 155){\large \sf(d)}
  }
  \vspace{-1.5cm}
  \centerline{
    \includegraphics[width=0.75\textwidth, angle=-90]{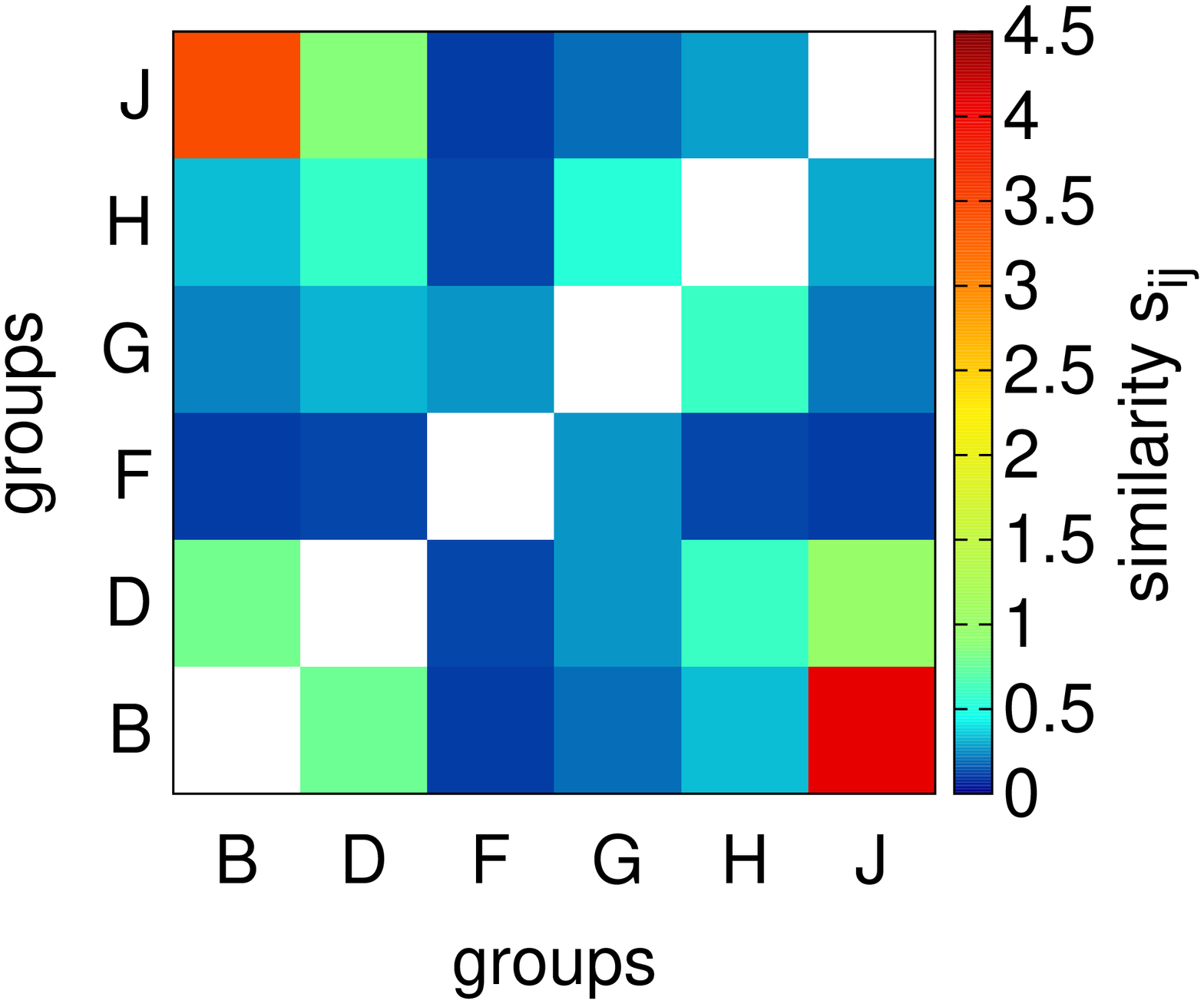}
    \put(-235, -40){\large \sf(e)}
    }
\end{minipage} 
\caption{\sf \textbf{Quantifying inter- and intra- group differences using BOC.}
\label{v-coef-fig}  (a) Numbers of calls in ensembles for pairwise group comparisons.
Inter group differences are larger than intra- group differences if the coefficients $v_{ij}$ (c) and $w_{ij}$ (b) are positive.
Confidence Intervals (orange shaded areas) are estimated by comparing randomly generated, not group-specific ensembles and assuming that the resulting values follow Student's distribution.
To test whether the similarity of group B and J is due to recordings made during the common encounter of group B and J, we repeated the analysis excluding all calls of group J that were recorded during the common encounter (green triangles).
The inverse of the weighted coefficients $s_{ij}=1/w_{ij}$ is expressed as the distance between edges in the network (d), and as colour coding of the network's adjacency matrix (e). 
}
\end{figure*}

\begin{figure*}[ht!]
  \vspace{-3cm}
  \begin{minipage}[ht!]{0.45\textwidth}
\centerline{
\includegraphics[width=1\textwidth]{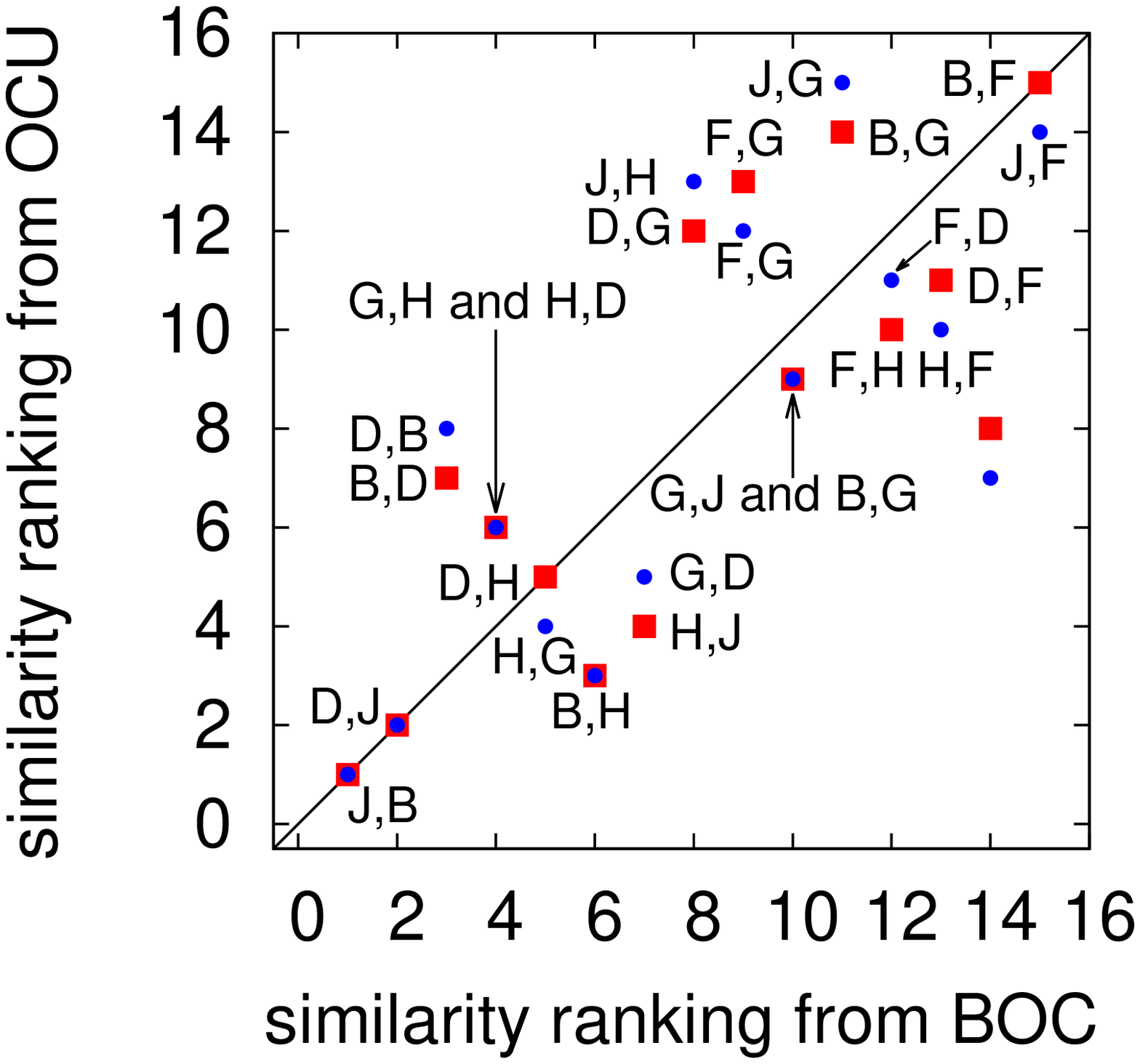}
\put(-160, 225){\large \sf(a)}
}
\end{minipage}
  \hspace{1.2cm}
  \begin{minipage}[t!]{0.4\textwidth}
    \vspace{-0.2cm}
\centerline{
\includegraphics[width=1\textwidth, angle=-90]{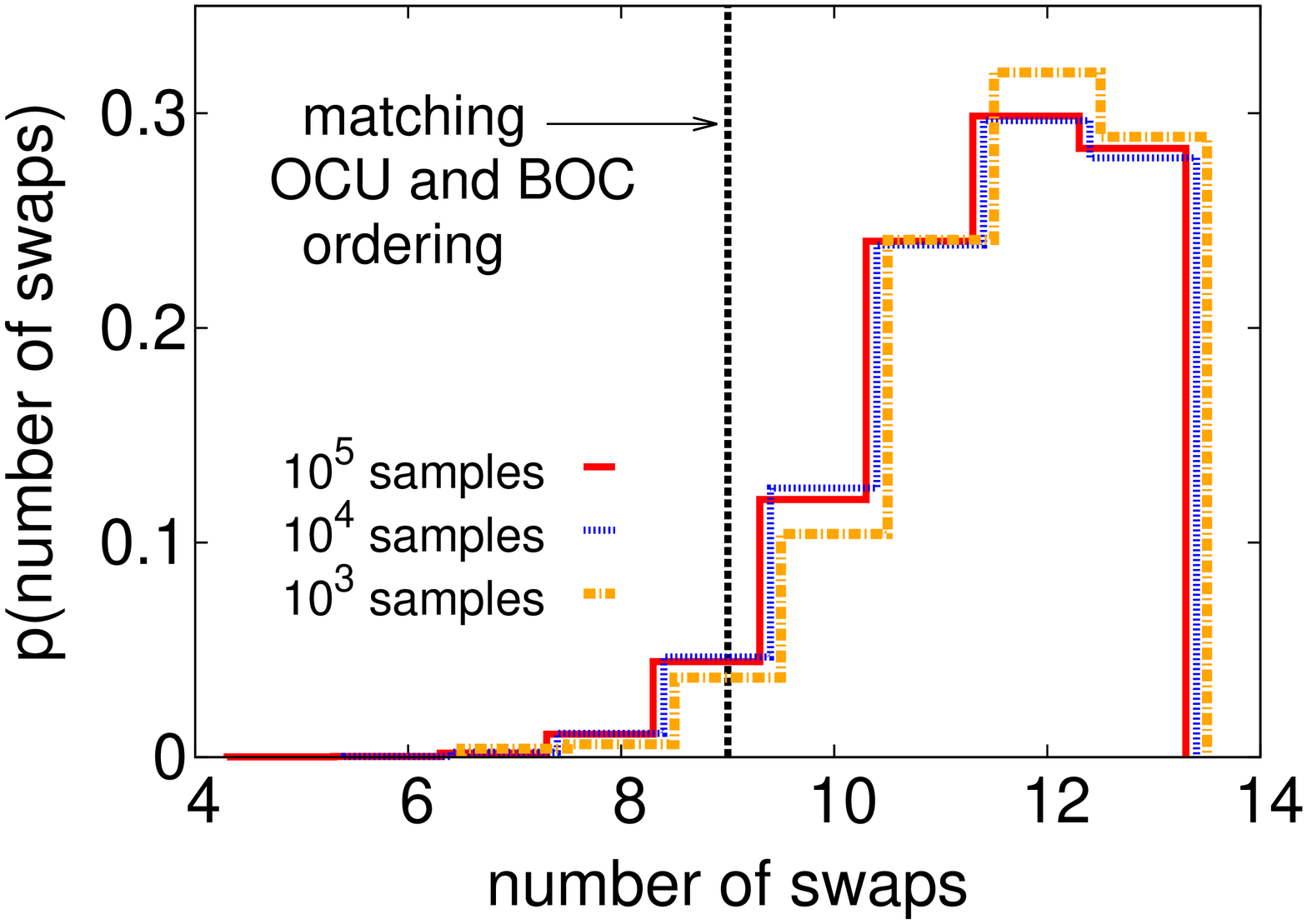}
\put(-223, -28){\large \sf(b)}
}
  \end{minipage}
  \vspace{-2cm}
\caption{\label{network-comparison} \sf \textbf{Comparing the results of OCU and BOC} (a) Ranking pairs of groups according to their similarity allows to compare the OCU (y-axis) and the BOC (x-axis) approach. Points on the diagonal represent pairs of groups which have the same similarity ranking according to both approaches. Two results (red squares and blue points) are obtained due to the asymmetry of the BOC approach with respect to the group that is chosen to estimate the intra-group similarity.
(b) The minimal number of swaps needed to transform the ranking according to the OCU approach into the BOC-ranking is $9$.
Comparing this number to the number of swaps needed for sorting random vectors of $15$ different integers (see distributions of $10^{3}$, $10^{4}$, and $10^{5}$ samples), we find that BOC- and OCU ranking are more similar to each other than $94$\% of all randomly created rankings.
}
\end{figure*}

\noindent{\sf \textbf{Overlap in call usage (OCU).}}
The six different recorded groups of pilot whales (group B, D, F, G, H and J) vocalised between seven (group F) and 54 (group H) different call types.
Fig.~\ref{groupvscalls} illustrates the usage of call types among different groups of whales, based on all calls classified into 140 call types.
Overlap in call usage (OCU) (see Fig.~\ref{groupvscalls}) between two groups was most common (30 call types), seven call types were shared among three groups and only one call type was shared between four groups.
Group B and group J had the highest call type overlap (N=12), group F and group G, as well as group G and group J had only one call type overlap.
Using the OCU as a measure for similarity in vocalisations we can later compare it to the similarity of groups obtained through the BOC-approach (see Figs.~\ref{v-coef-fig} and \ref{network-comparison}).

\noindent{\sf \textbf{Quantifying group-specificity with the bag-of-calls approach (BOC).}}
%
%
Using the BOC approach, we estimated group-specificity on the basis of ensembles ({\sl bags}) of calls.
In total, we used $1056$ non-overlapping calls, selected out of the $4582$ calls according to the quality of the sound files (high signal to noise ratio, no boat noise).
Note that this selection of mostly non-consecutive calls (31 min. length in total) is conceptually similar to a (random) sampling from all available recordings (17:32 h).
For each ensemble $E_{i}(n)$ representing group $i$, we compute a $128$ dimensional time series of cepstral coefficients and then estimate the distributions of these sound features.
We then quantify inter- and intra- group differences by computing the newly introduced group difference coefficients $v_{ij}$ and $w_{ij}$, (see Fig.~\ref{v-coef-fig}).
As shown in Fig.~(\ref{v-coef-fig}), $v_{ij}$ and $w_{ij}$ are larger than zero, indicating that inter- group differences are larger than intra group differences, for all pairwise comparisons. 
Each coefficient $v_{ij}$ and $w_{ij}$ can be evaluated in two different ways: with respect to the intra-group difference of the group $i$ and with respect to the intra-group difference of the group $j$.
Note that the number of calls in the respective ensembles $n =  \lfloor \frac{1}{2} \mbox{min}(N_i, N_j) \rfloor$ is the same for inter- and intra group comparisons.

For all but one out of $15$ group comparison (group B and J), the values of $v_{ij}$ and $w_{ij}$ are also clearly larger than the confidence bounds estimated using random ensembles of sizes n.
Since group B has been only observed while travelling and milling with group J, we tested whether the measured similarity in vocalisation could be influenced by the fact that the recordings were made at the same location.
Therefore, we repeated the comparison of group B and group J but excluded all recordings of group J that were made during the common encounter of group B and J (03/07/2010).
These additional results are shown as two single points (green triangles) in Fig.~\ref{v-coef-fig} and they are in the same order of magnitude as the previous comparison of group J and B.
Consequently we conclude that features of the vocalisations of group B and J are very similar, as far as we can estimate on the basis of cepstral distributions.

Using the inverse of the weighted coefficients $s_{ij}=1/w_{ij}$ as a measure for the similarity in vocalisation, we visualised the results of all comparisons in terms of a network with $s_{ij}$ being the adjacency matrix (see Fig.~\ref{v-coef-fig}). 
The distance between edges in this network corresponds to the values of $s_{ij}$, e.g., a high similarity in vocalisations of group B and J is represented by a small distance between edge B and J.
%
%

\noindent \textbf{Comparing results from BOC and the conventional OCU approach}
Comparing the adjacency matrices obtained through OCU (Fig.~\ref{groupvscalls} (c)) and BOC (Fig.~\ref{v-coef-fig} (e) ) one can see that the structures of both matrices are very similar.
To quantify this, we ranked all pairs of groups by their similarities as estimated by OCU and BOC (see Fig.~\ref{network-comparison}(a)), i.e. starting with the most similar groups (J,B) on rank 1 and finishing with the least similar groups (B, F) on rank 15.
The most similar groups (J,B) and (D,J) were identified by both methods, independently of the group that was used as a reference for intra-group similarity (blue circle and red square).
Also the similarity of groups (D,H) and (B,F) was ranked equally by OCU and BOC.
All other group comparisons were at least close to the diagonal which indicates a similar ranking by both methods. 
The minimal number of swaps needed to transform the ranking of the BOC-approach to the OCU ranking (9) is smaller than the average number of swaps needed to sort random vectors of $15$ integers into a specific ranking sequence (see Fig.~\ref{network-comparison} (b)).
Moreover, we find that BOC- and OCU ranking are more similar to each other than 94\% of all randomly created rankings.
Consequently, the BOC-approach produced results that are comparable to the observer based OCU of $4582$ calls, although only $1056$ calls of high sound quality were included in the BOC aproach.
%
%
\section*{\sf \textbf{Discussion}}
\vspace{-0.2cm}
We discover a complex group-specific vocal repertoire of discrete calls in the northern Norwegian pilot whale population.
This repertoire includes $140$ different discrete call types, several repetitive combinations of calls and whistles above 20 kHz.
Similar to the description of the vocal repertoire of long-finned pilot whales in the northwest Atlantic by Nemiroff and Whitehead \citep{Nemiroff2009} the repertoire of long-finned pilot whales in the northeast Atlantic comprise a variety of clicks and buzzes, low frequency calls that are noisy and irregular, as well as different types of whistles.
Additionally we find different types of pulsed calls that range from simple single-element calls to calls of highly complex structures composed of up to eight elements.
 Most pulsed calls consisted of only one element (93\% in Nemiroff \& Whitehead versus 86\% in our study). 
In addition to the higher number of pulsed calls with several elements, we also found more bi-phonal calls than in Nemiroff's study. 
Nemiroff and colleagues did not describe discrete patterns in the structure of their pulsed calls, whereas we had no problems identifying discrete calls and categorising them.
In addition to the smaller sample size of Nemiroff's study ($419$ calls), the different findings of this contribution (based on $4582$ calls) could be attributed to structural differences in the repertoire of northwest and northeast Atlantic long-finned pilot whales, or different methodological approaches to analyse and describe vocal repertoires.
These contrasting findings emphasise the importance of analysing vocal repertoires with different methodological approaches.

%
%
Surprisingly we also found ultrasonic whistles with frequencies above 20 kHz in the northern Norwegian long-finned pilot whale population, which were previously not described for this species.
Killer whales in Norway are known to produce whistles with ultrasonic frequencies (ranging up to 75 kHz) \citep{Samarra2010}.
The whistles we found for the long-finned pilot whales were similar in frequency range but different in structure and length. 
The reason for ultrasonic signals in top predators is unknown but it may be used in short range communication as has been suggested for killer whales \citep{Samarra2010}.

In general, the usage of whistles has been found to be highly context dependent\citep{Weilgart1990}: simple structured whistles are produced more often during resting behaviour such as milling, whereas more complicated structured whistles and pulsed calls occurred more frequently during active surface behaviour (probably including feeding behaviour). 
%
%
%
It has been suggested by Sayigh \citep{Sayigh2013} that short-finned pilot whales may use stereo-typed individual whistles, similar to the Bottlenose dolphin's signature whistles. 
%
%
We have observed Atlantic white-sided dolphins ({\sl Lagenorhynchus acutus}) travelling and foraging together with long-finned pilot whales;
during these inter-species encounters the whistle rates of the pilot whales increased.
Consequently it is possible that whistles are important during encounters with other species.
%

Temporal organisation in call sequences, as we observed for pilot whales, has rarely been reported and up to this day seems a rather rare event outside our own species. 
Arnold and Zuberb{\"u}hler published an example of call sequences in male putty-nosed monkeys ({\sl C. nictitans martini}) \citep{Arnold2006, Arnold2008}. 
Fitch and Hauser \citep{Fitch2004} tested cotton-top tamarins ({\sl Saguinus Oedipus}) with different sound sequences. 
Their results showed that the monkeys are capable of discriminating context-free grammars from finite state grammars, but not the other way around. 
In a recent study Alves and colleagues \citep{Alves2014} found that long-finned pilot whales match artificial sounds (sonar signals), supporting the view that the flexibility in vocal production of long-finned pilot whales 
is similar to the vocal flexibility of killer whales \citep{Ford1991, Miller2006}. 
In summary, long-finned pilot whales communicate with a high variety of vocal signals, social complexity is high, and their cognitive abilities range from vocal learning to mimicry and they show sound production flexibility.
%

%
Many automatised approaches to analysing whale vocalisations focus on the automated categorisation and classification of single sounds, such as types of calls and whistles \citep{Spong1993, Deecke1999, Deecke2005, Brown2007, Shapiro2009, Kaufman2012}.
Group-specific usage of vocalisations is then discussed by comparing the repertoire (which sounds are used) and whether there are variations of sounds.
In the first part of this contribution we follow this well established approach by conducting an observer based categorisation and classification of all recorded sounds.

To study group-dependent difference in vocalisation, we propose and test a new automated method, the BOC approach.
The main idea of this approach is that we omit separating and sorting vocalisations into call types, but compare ensembles of vocalisations produced by each group.
Investigating ensembles of calls rather than identifying individual call types is conceptually similar to the bag-of-words-model \citep{harris1954} used in text analysis.
In the original bag-of-words-model a text is represented as the {\sl bag} (multi-set) of words disregarding grammar and even word order but keeping multiplicity.
We investigate group-specific vocalisations by comparing ensembles i.e., {\sl bags} of-calls that contain calls of a specific group of whales.
Comparing the statistical properties of all features computed for each ensemble circumvents the necessity to establish subjective vocal categories or select specific acoustic features.
Conceptually similar, using histograms of sound features, has been suggested to attribute bird songs to bird species \citep{briggs2009}. 
Note that the way the ensembles of calls are constructed (choosing only high quality sounds and additionally applying a random sampling to data from several recording sessions per group) implies that calls within an ensemble most likely originate from different behavioural contexts and that the temporal correlation of calls is destroyed due to the random sampling.
Additionally, we did not select specific (hand-crafted) features, but use all information contained in the cepstral coefficients \citep{Bogert1963, briggs2009} of sounds.
Furthermore, we can even neglect the temporal ordering of these features and each group is well represented by their statistical distributions estimated for each ensemble of sounds.
We then quantified differences in vocalisation among six groups of pilot whales by computing differences in distribution.
To reason whether the calculated differences in distribution were relevant, we introduced two types of coefficients that summarise the relation between inter- and intra-group differences.

Inter-group differences were significantly larger then intra-group differences, for all but one out of $15$ inter-group comparisons.
Interestingly, groups B and J, the two groups with no significant difference in vocalisations, have also been observed travelling and milling together.
One possible explanation for their similarity in vocalisation is that they are related or that they are sub-groups of a bigger group.
The common encounter of group B and J allowed us also to estimate the effect that a similar acoustic environment could have on the similarity of two groups: even if calls recorded from group J during the common encounter are excluded from the analysis, we still find the same results when comparing ensembles of calls from group B and J. 
%
%
Since the calls of group J used for this later comparison were recorded on a different day at a different location, we can conclude that the effect of the different acoustic environments on the computed similarity of vocalisations is rather negligible.

Both, observer based classification of calls and the bag-of-calls approach yield similar results concerning group-specificity. 
Ranking pairs of groups according to their similarity, the BOC approach mostly reproduced the ordering of OCU approach which relies on observer based classification. 
This is surprising, since the BOC approach used less than 25\% of the number of calls which were considered in the OCU analysis.
Consequently, it is possible to distinguish groups of pilot whales automatically, by simply comparing ensembles of calls without referring to individual properties of single calls or comparing single call occurrences.

The correlation between high call type overlap (OCU results) and the overall similarity of the group’s vocal repertoire is a strong hint for vocal learning in this species.
Although all groups shared some call types, other call types were exclusively used by only one group. 
If calls are copied and learned when groups have contact with each other (either social or relatedness), then certain call types from other groups can be learned and integrated into their own repertoire.
With copying errors and innovation this can result over time into an overall similar call type repertoire of groups with high social contact and or close relatedness.

Group-specific vocalisation may be advantageous in several ways:
In the aquatic environment that favours matrilineal social organisation among mammals, it becomes important to recognise group members for offspring care, protection against predators and cooperative social and feeding behaviour. 
For roving males in those societies, group-specific vocal repertoires may help to distinguish between relatives and non-relatives to avoid inbreeding and to increase their fitness. 
In case of female choice, they will be able to distinguish between related and non-related mates by their vocal dialect.
However the concordance between relatedness and vocal similarity cannot be finally answered without genetic studies,

Setting our findings in relation to what is known about other tooth whales, killer whales in several places of the world are known to have dialects; first discovered in Canada in a resident population where groups use seven to 17 distinct call types \citep{Ford1991}). 
These repertoires remain stable over time, but subtle changes could occur in call type structure supporting the fact that killer whales have a high vocal flexibility \citep{Ford1991,Deecke2000}. 
Comparing the degree of shared vocalisation does reflect kinship in killer whale call types and sperm whale codas \citep{Deecke2010, Rendell2012}. 
The possibility to share more or less vocal types seems to be evolved via vocal learning, as social systems became more complex and the complexity of signals increased to recognise individuals, kin, or other social partners \citep{Janik1997}. 

Group-specific vocalisations are also known from other mammal societies, but the mechanism behind the specification seem to be different. 
Studies in nonhuman primates, gibbons and leaf monkeys, combining genetic and acoustic analysis of the same groups have shown a high concordance between genetic relatedness and acoustic similarity of their loud calls \citep{Thinh2012, Meyer2012}. 
But their loud calls are innate and no vocal learning is involved. In addition the degree of difference seems to be more influenced by sexual selection and not by social organisation. 
This is also true for some subtle group differences found in vocal repertoires of other nonhuman primates, like macaques \citep{Fischer1998} or chimpanzees \citep{Crockford2004}. 
In summary, we show that long-finned pilot whales in northern Norway exhibit a complex vocal repertoire both in call structure and variety of sounds. 
Proposing and testing a new automated method, we found a significant group-specificity of their vocalisations, hinting on the existence of vocal learning.
Additionally, we observed pilot whales to be flexible in the usage of calls and whistles, particularly when communicating with other groups and other species, which is reflected in call type sharing and similarity of the overall call properties.
Such high flexibility in vocal communication combined with vocal learning has so far only been reported for very few species.
%
%
\section*{\sf \textbf{Methods}}
\vspace{-0.2cm}
\subsection*{\sf \textbf{Ethics statement}}
\vspace{-0.2cm}
All observations and recordings reported in this contribution were made in the Vestfjord in northern Norway (see GPS coordinates in Tab.~\ref{tab:encounters}).
In general no permission is needed for non-invasive research on marine mammal along the Norwegian coast. 
To ensure that we conducted our research according to Norwegian ethical laws, we asked the Animal Test Committee (Forsøksdyrutvalget) of Norway for permission, and they confirmed that our studies do not require any permission (approval paper ID 6516).
%

\subsection*{\sf \textbf{Sound recording}}
\vspace{-0.2cm}
We used one or two Reson TC4032 hydrophones (frequency response 5 Hz - 120 kHz, omnidirectional), which were lowered directly ca.~ 18 m into the water from a 7 m Zodiac boat, when in close proximity (less than 50 m) to the whales.
Sound was amplified with a custom built Etec amplifier (DK) and recorded with different mobile recording devices; in 2006-2008 we used an Edirol-R09 (Roland) with a sampling frequency of 48 kHz, and in 2009-2010 we used a Korg MR-1000 with a sampling frequency of 192 kHz. 
GPS coordinates were taken of the beginning and end of an encounter, and notes of the whales' behaviour were continuously taken during recordings. 
Recordings lasted as long as the whales were within a 500 m range of the boat, as soon as they moved out of range and the signals became weak, we stopped the recordings and moved closer to the whales. 
At first sign of disturbance of the whales, we ceased the studies and waited 30 min before resuming our studies. 
If the whales were repeatedly disturbed we terminated the field encounter. 
In most cases, however the whales became quickly habituated to the presence of our boat and data collection was possible for longer periods.

%
\subsection*{\sf \textbf{Preparing bags-of-calls for intra- and inter-group comparisons}}
\label{ensembles}
\vspace{-0.2cm}
%
Basis of the bag-of-calls analysis are short recordings that have been cut from continuous data, such that each of them contains one call of a long-finned pilot whale.  
To test whether vocalisations are group-speciﬁc we only use call types of very good quality. 
%
%
Vocalisations of this quality originated only from utterances close to the boat and therefore ensure that only group members uttered these calls. 
To achieve a randomised sample of different group activities and group members, we use long recordings per group (ranging from 23 min to 5:14 h, mean: 2:55 h).
In total we include 1056 calls with sufficient quality in the analysis, with duration of the cut recordings varying between $0.14$ s and $6.27$ s.
\begin{table*}[ht!]
\sf
\caption{\sf \textbf{Numbers of calls $n$ used to create ensembles for inter- and intra-group comparisons}}
\hspace{1cm}
\begin{tabular}{ c c c c c c c }
\textbf{group} & \textbf{B} & \textbf{D} & \textbf{F} & \textbf{G} & \textbf{H} & \textbf{J} \\
\hline
\textbf{B} & 27, 45, 79, 84, 110 & 79 & 27 & 45 & 84 & 110\\ 
\hline
\textbf{D} & 79 & 27, 45, 79 & 27 & 45 & 79 & 79\\ \hline
\textbf{F} & 27 & 27 & 27 & 27 & 27 & 27\\ \hline
\textbf{G} & 45 & 45 & 27 & 27, 45& 45 & 45\\ \hline
\textbf{H} & 84 & 79 & 27 & 45 & 27, 45, 79, 84 &84\\ 
\hline
\textbf{J} & 110 & 79 & 27 & 45& 84& 27, 45, 79, 84, 110\\
\hline
\hline
\textbf{total \#} & {\multirow{2}{*}{221}} & {\multirow{2}{*}{159}} & {\multirow{2}{*}{54}} & {\multirow{2}{*}{90}} & {\multirow{2}{*}{168}} & {\multirow{2}{*}{337}}\\
\textbf{of calls} $N$ & & & & & & \\
\hline
\end{tabular}
\label{tab:ofn}
\end{table*}
%
The total length of all recordings used for this intra- and inter- group comparison was $31$ min. and $23$ s.
All recordings used in this part of the study have a sampling interval $\Delta t = 1/48000$ s. 
%
%

To compare group-specific differences in vocalisations, we construct several ensembles of calls ({\sl bags-of-calls}), each representing a sample of recordings made for one group of whales.
The detailed procedure of constructing bags-of-calls is described as follows:
Suppose we have $N_i$ recordings of calls of whale group $i$ and $N_j$ recordings of group $j$.
An ensemble $E_i(n)$ of $n$ calls is drawn through random number generators out of all calls that have been recorded for group $i$, with $i=B,D,F,G,H,J$. 
We then randomly draw a second ensemble $E_j(n)$ consisting of $n$ out of $N_j$ recordings of the second group $j$, with $j\neq i$. 
For each pairwise comparison of groups, the size of the ensembles $n$ is adjusted according to the smaller number of available recordings, i.e., to the largest integer smaller than half of the smaller number of available recordings, $n = \lfloor \frac{1}{2} \mbox{min}(N_i, N_j) \rfloor$.
Considering e.g., group $J$ and $B$, there were $337$ recordings of group $J$ and $221$ recordings of group $B$ available. 
Consequently the size of the ensembles generated for comparing groups $J$ and $B$ should e.g., be $n=110$.
We then compare group $i$ and $j$ by comparing properties of the ensembles $E_{i}(n)$ and $E_{j}(n)$. 
In order to check whether the resulting differences can be attributed to the difference in group, we also compare pairs of two random ensembles drawn from the same group,
using the $N_{i}-n$ or $N_{j}-n$ remaining recordings (i.e., recordings not used for the previous comparison of two different groups).
In other words, we generate additional ensembles $\tilde{E}_{i}(n)$ of size $n$ using now only the recordings that are {\sl not} part of ensemble $E_i(n)$. 
The differences in distribution within group $i$ are then estimated by comparing properties of $E_{i}(n)$ and $\tilde{E}_{i}(n)$. 
%

%
\subsection*{\sf \textbf{Comparing distributions of features}}
\vspace{-0.2cm}
\label{comparing-distributions}
For each ensemble $E_{i}(n)$ representing group $i$, we compute time series of sound features and then estimate the distributions of these features.
In more detail, we compute time series of the $q$-th cepstral coefficient $c_{q,t}$, with $t=0, ..., T$ and $T$ being the total length of all calls within the ensemble, with $q=1,2, ..., 128$ and then estimate the distributions $p_{q,i}$ of each data set.
Neglecting the time dependencies within each call and within the ensemble, we only consider the collection of all features within a given ensemble. 
%
\begin{figure}[ht!]
\centerline{
\includegraphics[width=0.28\textwidth]{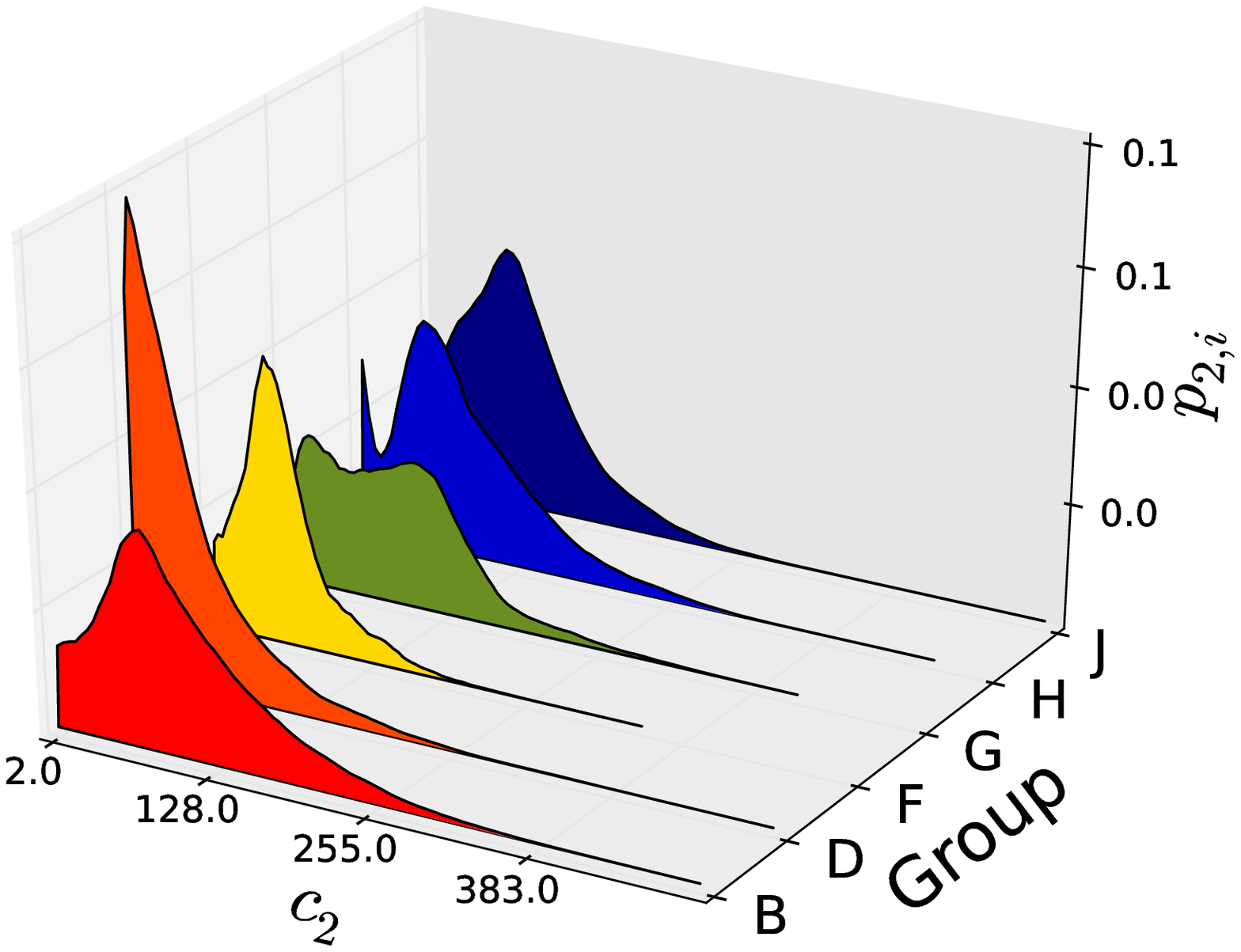}
\put(-120,90){(a)}
\hspace{-1cm}
\includegraphics[width=0.28\textwidth]{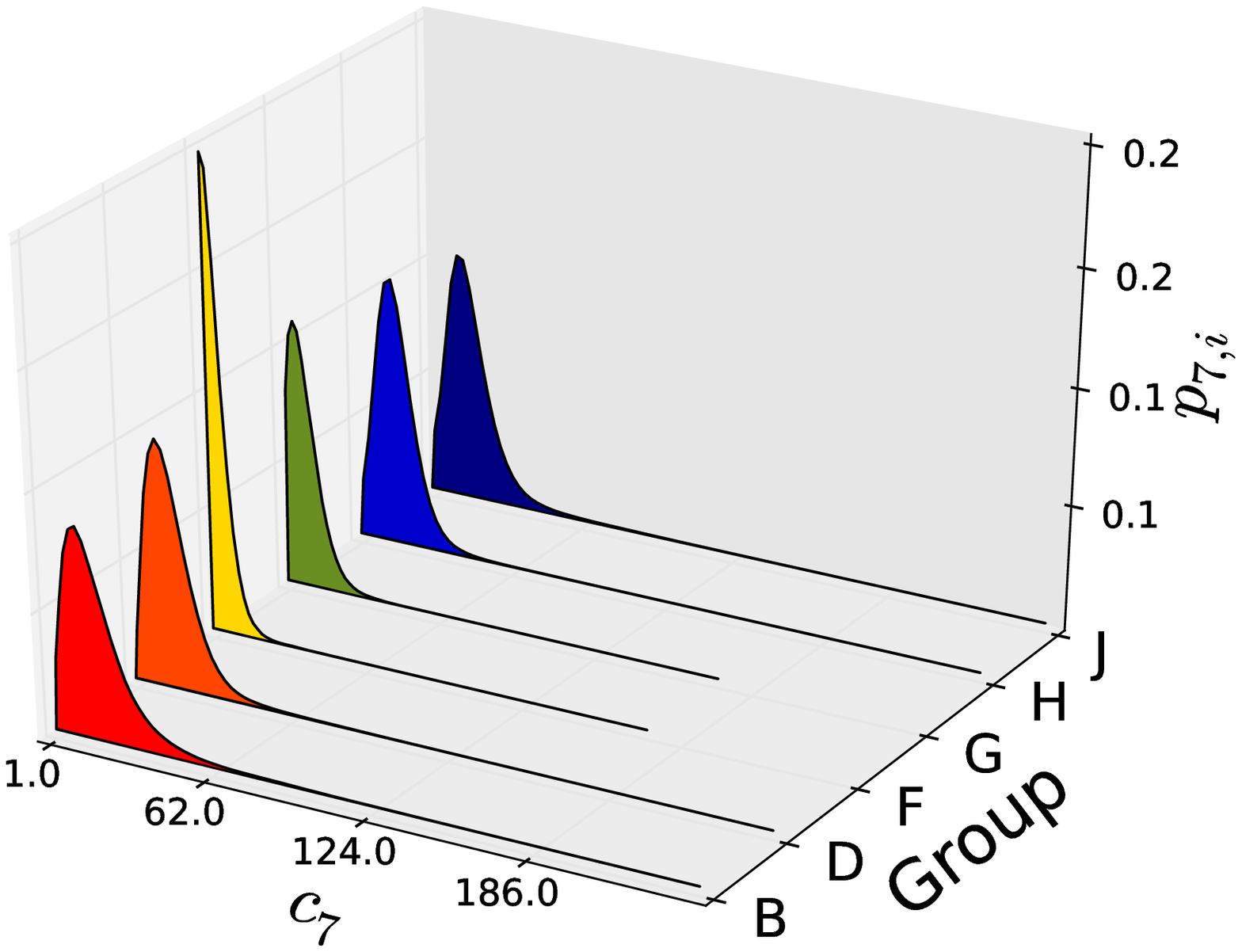}
\put(-120,90){(b)}
}
\vspace{-0.3cm}
\centerline{
\includegraphics[width=0.28\textwidth]{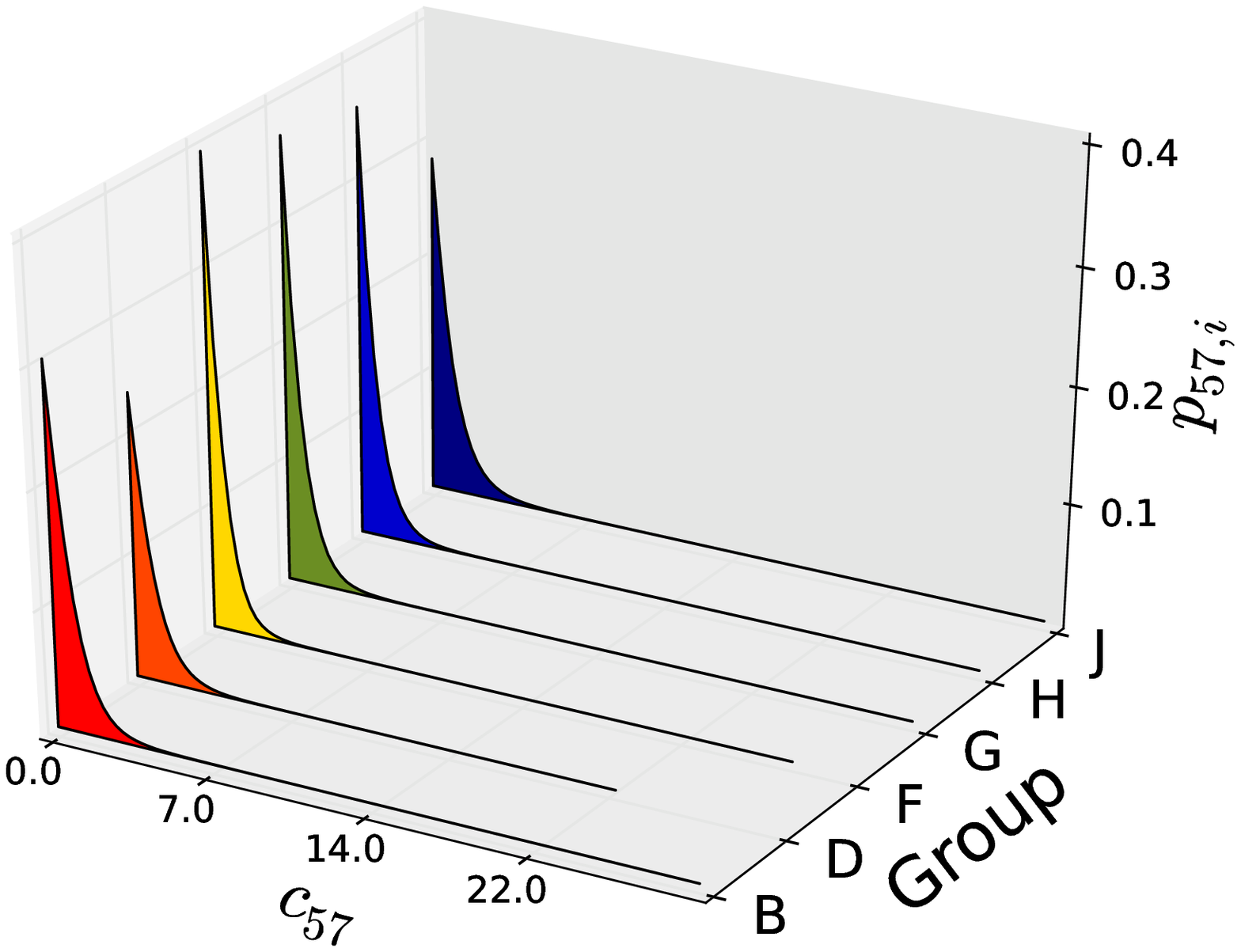}
\put(-120,90){(c)}
\hspace{-1cm}
\includegraphics[width=0.28\textwidth]{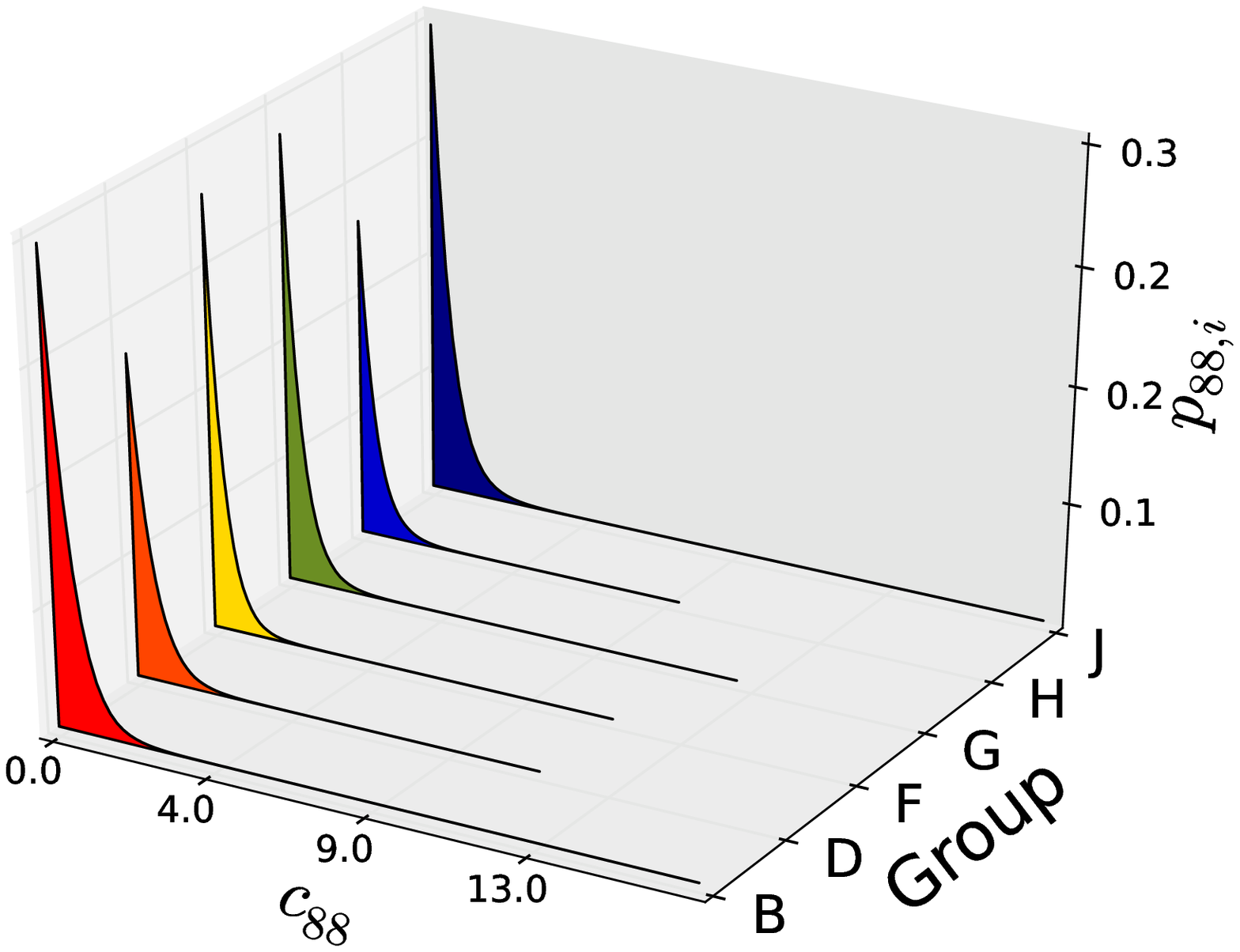}
\put(-120,90){(d)}
}
\caption{\label{dist-example} \sf \textbf{Four examples for distributions of cepstral coefficients} $c_{n,t}$ : (a), $p_{2,i}$ (b), $p_{6,i}$, (c) $p_{57,i}$ and (d) $p_{88,i}$ estimated for groups $i = \{ B, D, F, G, H, I\}$. The number of calls used to estimate these distributions are chosen to be the largest number used for ensemble comparisons, as specified in Tab.~\ref{tab:ofn}.
}
\end{figure}
We thus propose that properties of the ensemble $E_i(n)$ are represented by the distribution $p_{q,i}$ of the values of each coefficient $c_{q,t}$ of group $i$.
We then ask in how far the $128$ distributions (one for each cepstral coefficient $c_{q,t}$, $q=1,2,..., w/4$) for ensemble $E_i(n)$ are different from the distributions of another ensemble $E_j(n)$.
Fig.~\ref{dist-example} shows four examples for distributions of cepstral coefficients estimated for ensembles of calls recorded from different groups of whales.
For some coefficients, in particular for coefficients with smaller $q$, [e.g., Fig.~\ref{dist-example} (a)] the differences between distributions are even noticeable by visual inspection,
whereas, many higher order coefficients, representing small scale fluctuations in the spectrum have more similar distributions.

To quantify this observation, we use common measures for the difference in distribution, such as the (symmetrised) Kullback Leibler divergence  $D(p_{q,i}|| p_{q,j})$ \citep{KL1951} or the Hellinger distance \citep{Hellinger1909}.
%
%
%
%
%
Here, we use a symmetrised version of the Kullback Leibler divergence 
\begin{equation}
S(p_{q,i} || p_{q,j}) = D(p_{q,i}|| p_{q,j}) + D(p_{q,j}|| p_{q,i})\quad, \label{skld}
\end{equation}
to quantify the difference between distributions $p_{q,i}$ and $p_{q,j}$ of the $q$-th coefficient, representing group $i$ and group $j$.
Similarly one can compute $S(p_{q,i} || \tilde{p}_{q,i})$ to measure intra- group differences by comparing the distributions $p_{q,i}$ and $\tilde{p}_{q,i}$ referring to ensembles of equal size drawn from recordings of the same group $i$, as explained above.

To summarise and quantify the difference between inter- and intra- group comparisons we introduce coefficients
\begin{equation}
v_{ij} = \sum_{q}  \left[ S(p_{q,i} || p_{q,j}) - S(p_{q,i} || \tilde{p}_{q,i}) \right] \quad. \label{v-equation}
\end{equation}
%
%
Inter- group differences are larger than intra- group differences is $v_{ij}$ is positive and vice versa if $v_{ij}$ is negative.
Since smaller coefficients reflect large scale structures in spectra, differences in distributions of small coefficients can be considered to be more relevant than differences in higher order coefficients that can be due to small scale fluctuations and noise.
Therefore, we additionally introduce linearly weighted coefficients
\begin{equation}
w_{ij} = \sum_{q}  \frac{1}{q}\left[ S(p_{q,i} || p_{q,j}) - S(p_{q,i} || \tilde{p}_{q,i}) \right] \quad, \label{w-equation}
\end{equation} 
%
as a second measure for comparing inter- and intra-group differences.
The relevance of the coefficients with smaller index $q$ is emphasised trough larger weights $1/q$.

To test whether the calculated values of $v_{ij}$ and $w_{ij}$ are statistically significant we estimate confidence intervals for both coefficients by comparing randomly generated ensembles $E_{r}(n)$.
In more detail, we construct $m$ such ensembles by randomly drawing $n$ recordings out of all available recordings (not sorted by group) without repetition.
Note that these mixed random ensembles used to estimate confidence intervals contain recordings from different groups.
Any $3$-tuple of random ensembles, $E_{r}$, $E_{r'}$, and $E_{r''}$ can serve to simulate a comparison of inter- and intra- group differences.
Consequently, two random ensembles, e.g. $E_{r}$ and $E_{r''}$ are interpreted as representing the same group, whereas the third one ($E_{r'}$) is assumed to represent a different group.
For each $3$-tuple of random ensembles, we compute the coefficients $v_{rr'}$, $v_{rr''}$ and $w_{rr'}$, $w_{rr''}$.
We then use the distributions of $g(m)=\sum_{k=0}^{m-1} a(k)$ values for each coefficient (with $a(k)$ referring to triangular numbers) to estimate confidence intervals according to Student's distribution.
Additionally, we introduce the inverse of the weighted coefficient, $s_{ij}=1/w_{ij}$ as a measure for the similarity in vocalization. 
%
%
\section*{\sf \textcolor{dg}{\textbf{Acknowledgments}}}
%

%
We thank Denny Fliegner, Theo Geisel, Jan Nagler and Julia Fischer for support during project initiation.
We also thank Andre Baumeister for providing a map of Vestfjorden and Fredrik Broms, Lotta Borg,
Kerstin Haller and Madita Zetzsche for field assistance and photo-identification. 
Special thanks also for Bogna Waters for proof-reading of the manuscript and
Patrick Kramer for creating a literature database of many relevant articles. 
This study was supported by Ocean Sounds, the World Wildlife Fund Sweden and
the Max Planck Society.
{\small
\bibliographystyle{naturemag_noURL}
\bibliography{pilot-whales-no-urls}
}
\end{document}